\documentclass[%
 reprint,
 amsmath,amssymb,
 aip,
 jap,
]{revtex4-1}

\usepackage{graphicx}% Include figure files
\usepackage{dcolumn}% Align table columns on decimal point
\usepackage{bm}% bold math
\usepackage{tikz}
\usetikzlibrary{decorations.pathreplacing}
\usepackage{subfigure}
\begin{document}

\preprint{APS/123-QED}

\title{A Modified and Calibrated Drift-Diffusion-Reaction Model for Time-Domain 
Analysis of Charging Phenomena in Electron-Beam Irradiated Insulators}% Force line breaks with \\

\affiliation{Delft Institute of Applied Mathematics, Delft University of Technology,
 Mekelweg 4, 2628 CD Delft, The Netherlands}
\author{Behrouz Raftari}
\author{Neil Budko}
\author{Kees Vuik}
\affiliation{Delft Institute of Applied Mathematics, Delft University of Technology,
 Mekelweg 4, 2628 CD Delft, The Netherlands}

\date{\today}% It is always \today, today,

\begin{abstract}
This paper presents an improved version of 
the previously proposed self-consistent drift-diffusion-reaction model correcting for
non-physical behavior at longer time scales. 
To this end a novel boundary condition is employed that takes into account 
the effect of tertiary electrons and a fully dynamic trap-assisted generation-recombination 
mechanism is implemented.
Sensitivity of the model with respect to material parameters is investigated
and a calibration procedure is developed that reproduces experimental yield-energy 
curves for uncharged insulators. Long-time charging  
and yield variations are analyzed for stationary defocused and focused beams as well as
moving beams dynamically scanning composite insulators. 

\end{abstract}

%Use showkeys class option if keyword
\maketitle

%%%%%%%%%%%%%%%%%%%%%%%%%%%%%%%%%%%%%%%%%%%%%%%%%%%%%%%%%%%%%
\section{\label{sec:level1}Introduction}
%%%%%%%%%%%%%%%%%%%%%%%%%%%%%%%%%%%%%%%%%%%%%%%%%%%%%%%%%%%%%
Charging phenomena in insulators have long been studied due to their importance in 
such areas as scanning electron microscopy (SEM), memory-based technologies, particle detectors, 
ceramic surfaces, industrial cables, and the safety of spacecraft \cite{lee2007layer,vdGraaf2017,
kulwicki1970diffusion,bass1998measurements,vampola1985aerospacer}.
Probably, the earliest systematic studies of electron-irradiation effects in solids and 
charging phenomena in insulators, as parts of research on electrets, 
were carried out by B.~Gross who has had a great impact on this 
research field. In his seminal works on irradiation phenomena
\cite{gross1957irradiation,gross1958irradiation} Gross investigated the electron trapping 
and charge buildup in high-resistivity solid insulators bombarded with 
energetic electrons. Further studies by Gross and coworkers produced
new experimental techniques and mathematical models 
\cite{gross1967high,gross1973charge,gross1974charge,gross1974transport}.

These and more recent \cite{liebl1980sims,le1989study,meyza2003secondary,
touzin2006electron,li2011self,li2010surface,walker2016simulations} studies have not 
yet been able to provide a complete and coherent account of all observed 
phenomena. This could be due to the prevailing emphasis on static (stationary) 
models\cite{dionne1975origin,barut1954mechanism,henke1979soft}
rather than time-domain analysis. 
Studying the dynamics of charging 
in time domain is especially important in the analysis of response times
in particle detectors\cite{vdGraaf2017} and in designing novel scanning strategies 
for SEM\cite{walker2016simulations}.
The existing dynamic 
models are either one-dimensional \cite{meyza2003secondary,touzin2006electron}
or do not include some of the essential physical processes, e.g., dynamic recombination,
trapping, etc\cite{li2011self,li2010surface}. 

While the prevailing semi-classical Monte-Carlo (MC) method \cite{kieft2008refinement}
makes very few assumptions about the complicated electron-sample interaction process,
realizing its full theoretical potential is technically very challenging.
First of all, MC simulations are slowed down by the need to continuously update the
long-rage electrostatic potential.
Secondly, it is computationally difficult to keep track of all the 
trapped and de-trapped electrons.
Finally, achieving acceptable variance not only in particle numbers, but also 
in the times of events (e.g. emission times), may require a prohibitive number of 
statistical realizations. 

Instead of sampling the probability space,
the drift-diffusion-reaction (DDR) approach, mainly used to model low-energy transport in 
semiconductors \cite{entner2007modeling,markowichsemiconductor}, 
directly describes the space-time evolution of a continuous probability density 
function.
The pertaining partial differential equations are obtained from the 
semi-classical Boltzmann equation applying the method of moments 
and a few assumptions about the distribution of particles over the momentum space.
From the mathematical point of view the DDR approach assumes 
that the symmetric part of the secondary electrons (SE) probability density function 
is isotropic about the origin of the momentum space and is 
well-described by a shifted Maxwellian distribution.

In our previous publication \cite{raftari2015self} we argued 
that the DDR approach can be applied to electron-beam 
irradiated insulators if the initial high-energy transport stage
is approximated by an empirical source function.
We showed that this pulsed source function allows modeling both the short-time 
processes immediately following the primary electron (PE) impact and the
long-time charge evolution due to sustained bombardment. Importantly, we demonstrated
that the sustained irradiation can also be modeled by a continuous current source,
which gives practically the same secondary electron (SE) emission current as the time-averaged 
SE emission produced by many single-impact pulsed sources.

However, the original DDR model \cite{raftari2015self} had serious 
shortcomings as well. First of all, it was not calibrated against experimental 
data. Although we were able to reproduce any SE yield at 
a chosen PE energy by tuning a single parameter -- the electron emission velocity 
(surface recombination velocity) at the sample-vacuum interface,
it was not clear which yield should be taken as a reference, since yields tend to change
over time and depend on beam currents. Secondly, using the same SE emission velocity for all PE energies
resulted in curves not fully compatible with published SE yield data over 
the whole range of PE energies.
And more seriously, the model produced nonphysical results in the case of 
prolonged irradiation. Namely, the surface potential at low PE energies could reach 
very large positive values, which is not possible, since positive potential
attracts secondary electrons back to the sample leading to the neutralization of
any potentials exceeding $\sim 10$~V. 

We have identified the main reasons behind the bad long-time behavior of the original 
DDR approach\cite{raftari2015self}. These were the employed steady-state 
generation-recombination model, which is not really suitable for the analysis of 
transient effects, and the neglected tertiary electron current. 
Incorporating fully dynamic generation and recombination processes
is relatively easy. Here we employ the so-called trap-assisted 
generation-recombination model, which also reduces the number of 
equations to be solved and charge species to be tracked.

In hybrid MC-DDR methods \cite{li2011self,li2010surface}
tertiary currents are estimated with direct MC 
simulations of particle trajectories. Here we propose an alternative
approach that keeps intact the self-consistent nature of the DDR model.
Namely, we introduce a novel boundary condition at the sample-vacuum 
interface that accounts not only for the total number of 
electrons returning to the sample, but also for the spatial distribution 
of this tertiary current along the sample interface. 

We have also developed and implemented a clear calibration procedure
for our DDR model. It uses the fact that certain types of yield measurements 
-- the so-called standard-yield measurements --
correspond to the situation where single PE impacts happen sufficiently 
far enough from each other across the sample surface for their mutual 
interaction to be neglected. As our code is able to simulate single 
impacts, its calibration can be performed in this single-impact mode.

In Section~\ref{sec:Model} we outline the mathematical details of the modified 
DDR model. Its physical applicability is further discussed in Section~\ref{sec:calibration}, 
where we also investigate the sensitivity of the model, explain our parameter choices, 
develop a calibration procedure against published experimental data, 
and compare our results for defocused beams with 
an alternative one-dimensional approach. 
Section~\ref{sec:focused} presents further quantitative analysis of
more realistic scenarios with
focused stationary and moving beams, including a dynamic 
line-scan of a laterally inhomogeneous target.

%%%%%%%%%%%%%%%%%%%%%%%%%%%%%%%%%%%%%%%%%%%%%%%%%%%%%%%%%%%%%
\section{\label{sec:Model}Modified DDR model}
%%%%%%%%%%%%%%%%%%%%%%%%%%%%%%%%%%%%%%%%%%%%%%%%%%%%%%%%%%%%%
In this section we recall the main features of the DDR model\cite{raftari2015self} 
and describe several significant modifications that have been implemented 
since its introduction. 

\subsection{Basic equations}
%%%%%%%%%%%%%%%%%%%%%%%%%%%%%%%%%%%%%%%%%%%%%%%%%%%%%%%%%%%%%
The continuum approximation of the equilibrium transport of charged particles 
in insulators consists of both partial (PDE) and ordinary (ODE) differential 
equations augmented with a semi-empirical source function accounting for 
the initial ballistic transport stage.
The PDE's are the Poisson equation for the potential and the transport equations for the 
free charge density:
    \begin{align}\label{equ:Po}
         -\nabla\cdot(\varepsilon\nabla V)=\frac{q}{\varepsilon_0}(C+p-n-n_T),
    \end{align}
    \begin{align}\label{equ:electron}
         \frac{\partial n}{\partial t}+\nabla\cdot\mathbf{J}_n=S_n-(R_n-G_n),
    \end{align}
    \begin{align}\label{equ:hole}
         \frac{\partial p}{\partial t}+\nabla\cdot\mathbf{J}_p=S_p-(R_p-G_p),
    \end{align}
with the constitutive relations for the current densities given by
\begin{align}\label{equ:elec-dens}
 \textbf{J}_n= -D_n\nabla n+\mu_nn\nabla V,
\end{align}
\begin{align}\label{equ:hole-dens}
 \textbf{J}_p=-D_p\nabla p-\mu_pp\nabla V,
\end{align}
where $q$ is the elementary charge, $V(\mathbf{x},t)$ is the electrostatic potential, 
$n(\mathbf{x},t)$ is the density of free electrons, $n_T(\mathbf{x},t)$ is the density 
of trapped electrons, $p(\mathbf{x},t)$ is the density of free holes, $C$ 
is the density of empty traps, the constant $\varepsilon_0$ is the dielectric constant 
of vacuum, the function $\varepsilon(\mathbf{x})$ is the (static) relative permittivity 
of the sample, $\mu_n$ and $\mu_p$ are the electron and hole mobilities and 
$D_n$ and $D_p$ are the diffusion coefficients.

\subsection{Generation, recombination, trapping, and de-trapping}
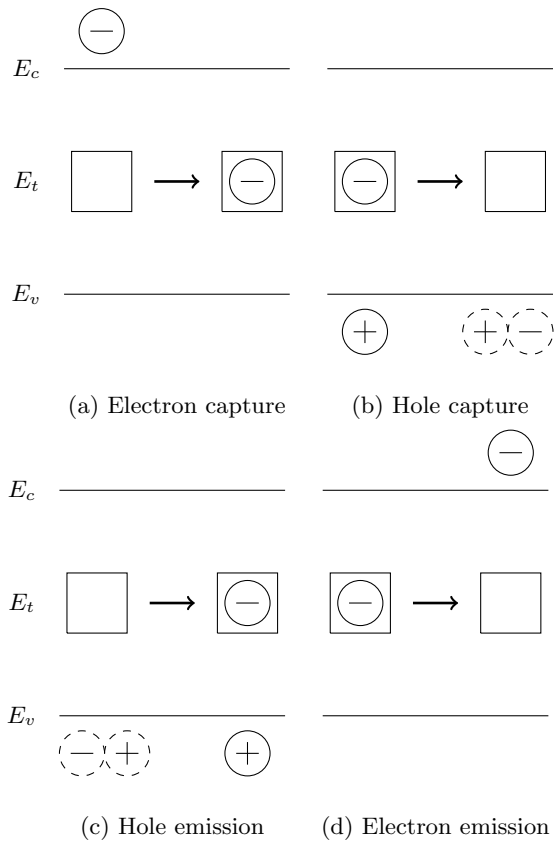
\begin{figure}[t!]
\begin{tikzpicture}
     % free electron
     \draw (-1,2) circle (0.3);
     \draw[-] (-1.15,2) -- (-0.85,2);
     % energy level
     \draw[-] (-1.5,1.5) -- (1.5,1.5);
     % arrow line
%      \draw[line width=1pt,->] (0.5,1.3) -- (1.5,0.8); %curved arrow
%      \node[] at (0.3,1) {Energy};
     % empty trap 
     \draw (-1.4,-0.4) -- (-0.6,-0.4) -- (-0.6,0.4) -- (-1.4,0.4) -- (-1.4,-0.4);
     % trapped electron
     \draw (0.6,-0.4) -- (1.4,-0.4) -- (1.4,0.4) -- (0.6,0.4) -- (0.6,-0.4);
     \draw (1,0) circle (0.3);
     \draw[-] (0.85,0) -- (1.15,0);
     % arrow line
     \draw[line width=1pt,->] (-0.3,0) -- (0.3,0);
     % energy level
     \draw[-] (-1.5,-1.5) -- (1.5,-1.5);
     \node[] at (-2,1.5) {$E_c$};
     \node[] at (-2,0) {$E_t$};
     \node[] at (-2,-1.5) {$E_v$};
     % caption
     \node[] at (0,-3) {(a) Electron capture};
%----------------------------------------------------------------
%----------------------------------------------------------------
     \draw[-] (2,1.5) -- (5,1.5);
     % arrow line
%      \draw[line width=1pt,->] (4,-1.4) -- (5,-1);
%      %
%      \node[] at (3.8,-1) {Energy};
     % empty trap
     \draw (4.1,-0.4) -- (4.9,-0.4) -- (4.9,0.4) -- (4.1,0.4) -- (4.1,-0.4);
     % arrow line
     \draw[line width=1pt,->] (3.2,0) -- (3.8,0);
     % trapped electron 
     \draw (2.1,-0.4) -- (2.9,-0.4) -- (2.9,0.4) -- (2.1,0.4) -- (2.1,-0.4);
     \draw (2.5,0) circle (0.3);
     \draw[-] (2.35,0) -- (2.65,0);
     % energy level line 
     \draw[-] (2,-1.5) -- (5,-1.5);
     \draw (2.5,-2) circle (0.3); % free hole
     \draw[-] (2.35,-2) -- (2.65,-2);
     \draw[-] (2.5,-2.15) -- (2.5,-1.85);
     \draw [dashed] (4.7,-2) circle (0.3); % dashed electron
     \draw[-] (4.55,-2) -- (4.85,-2);
     \draw [dashed] (4.1,-2) circle (0.3); % dashed hole
     \draw[-] (3.95,-2) -- (4.25,-2);
     \draw[-] (4.1,-2.15) -- (4.1,-1.85);
     \node[] at (3.5,-3) {(b) Hole capture};
\end{tikzpicture}
\begin{tikzpicture}
     % energy level
     \draw[-] (-1.5,1.5) -- (1.5,1.5);
     % arrow line
%      \draw[line width=1pt,->] (-1.5,-0.8) -- (-0.5,-1.3); %curved arrow
%      \node[] at (-0.3,-1) {Energy};
     % empty trap 
     \draw (-1.4,-0.4) -- (-0.6,-0.4) -- (-0.6,0.4) -- (-1.4,0.4) -- (-1.4,-0.4);
     % trapped electron
     \draw (0.6,-0.4) -- (1.4,-0.4) -- (1.4,0.4) -- (0.6,0.4) -- (0.6,-0.4);
     \draw (1,0) circle (0.3);
     \draw[-] (0.85,0) -- (1.15,0);
     % arrow line
     \draw[line width=1pt,->] (-0.3,0) -- (0.3,0);
     % energy level
     \draw[-] (-1.5,-1.5) -- (1.5,-1.5);
     \draw [dashed] (-1.2,-2) circle (0.3); % dashed electron
     \draw[-] (-1.05,-2) -- (-1.35,-2);
     \draw [dashed] (-0.6,-2) circle (0.3); % dashed hole
     \draw[-] (-0.45,-2) -- (-0.75,-2);
     \draw[-] (-0.6,-2.15) -- (-0.6,-1.85);
     \draw (1,-2) circle (0.3); % free hole
     \draw[-] (1.15,-2) -- (0.85,-2);
     \draw[-] (1,-2.15) -- (1,-1.85);
     \node[] at (0,-3) {(c) Hole emission};
     
     \node[] at (-2,1.5) {$E_c$};
     \node[] at (-2,0) {$E_t$};
     \node[] at (-2,-1.5) {$E_v$};
     %-------------------------------------------------------
     % free electron
     \draw (4.5,2) circle (0.3);
     \draw[-] (4.65,2) -- (4.35,2);
     \draw[-] (2,1.5) -- (5,1.5);
     % arrow line
%      \draw[line width=1pt,->] (2,1) -- (3,1.4);
%      %
%      \node[] at (3.2,1) {Energy};
     % empty trap
     \draw (4.1,-0.4) -- (4.9,-0.4) -- (4.9,0.4) -- (4.1,0.4) -- (4.1,-0.4);
     \draw[line width=1pt,->] (3.2,0) -- (3.8,0);
     % trapped electron 
     \draw (2.1,-0.4) -- (2.9,-0.4) -- (2.9,0.4) -- (2.1,0.4) -- (2.1,-0.4);
     \draw (2.5,0) circle (0.3);
     \draw[-] (2.35,0) -- (2.65,0);
     % energy level line 
     \draw[-] (2,-1.5) -- (5,-1.5);
     \node[] at (3.5,-3) {(d) Electron emission};
\end{tikzpicture}
\caption{Trap-assisted generation/recombination model.}
%---------------------------------------
\label{fig1}
\end{figure}

%%%%%%%%%%%%%%%%%%%%%%%%%%%%%%%%%%%%%%%%%%%%%%%%%%%%%%%%%%%%%
In the present modification of the DDR approach the dynamic trap-assisted Shockley-Read-Hall 
(SRH) generation/recombination model is implemented. 
Here we explain it along the lines of the PhD~study by Robert Entner  
conducted at TU Wien \cite{entner2007modeling}. An attractive feature of
this model is that there is no need to 
keep the track of trapped holes as all the relevant physics is already contained
in the single equation for the rate of electron trapping:
\begin{align}\label{equ:nTrap}
\frac{\partial n_T}{\partial t}= (R_n-G_n)-(R_p-G_p).
\end{align}
This process is coupled to the basic equations (\ref{equ:Po})--(\ref{equ:hole-dens})
and can be divided into four subprocesses illustrated in Fig.~\ref{fig1}.

(a)~{\it Electron capture}: An electron from the conduction band gets trapped at 
the band-gap of the insulator and the surplus energy of $E_c-E_t$ is transmitted 
to the phonon emission. The average rate of this process is
\begin{align}
 R_n=\sigma_n\upsilon_{th}n(N_T-n_T).
\end{align}

(b)~{\it Hole capture}: A trapped electron  
moves to the valence band and neutralizes a hole (i.e. the
hole is captured by the occupied trap), producing a phonon with the 
energy $E_t-E_v$. The corresponding rate is
\begin{align}
 R_p=\sigma_p\upsilon_{th}pn_T.
\end{align}

(c)~{\it Hole emission}: An electron leaves a hole in the valence band and is trapped 
(i.e. the hole is emitted from the empty trap to the valence band). The energy $E_t-E_v$ 
is needed for this process, and the corresponding rate is
\begin{align}
 G_p=\sigma_p\upsilon_{th}n_i(N_T-n_T).
\end{align}

(d)~{\it Electron emission}: A trapped electron moves to the conduction
band. The required energy is $E_c-E_t$, and the rate is 
\begin{align}
 G_n=\sigma_n\upsilon_{th}n_in_T.
\end{align}

%
%%%%%%%%%%%%%%%%%%%%%%%%%%%%%%%%%%%%%%%%%%%%%%%%%%%%%%%%%%%%%%%%%%%%%%%%%%%%%%%%%%%%%%%
%%%%%%%%%%%%%%%%%%%%%%%%%%%%%%%%%%%%%%%%%%%%%%%%%%%%%%%%%%%
%
In the above equations: $\sigma_n(\mathbf{x})$ and $\sigma_p(\mathbf{x})$ 
are the electron and hole mean trapping cross sections, 
$N_T(\mathbf{x})$ is the total density of traps, $n_T(\mathbf{x},t)$ is the density 
of trapped electrons, $\upsilon_{th}(\mathbf{x})$ is the thermal velocity, and 
$n_i(\mathbf{x})$ is the intrinsic carrier density. The spatial variable $\mathbf{x}$
indicates the possibility of spatial inhomogeneity, i.e., the presence of 
different adjacent materials.

The initial conditions on $n$ and $p$ at $t=0$ are set as the corresponding intrinsic 
carrier densities of the materials under consideration, whereas the initial condition 
for $n_T$ has been derived based on the assumption of the initial steady state for the 
density of trapped electrons prior to the start of irradiation 
(i.e. $\partial n_T/\partial t=0$) and is set to 
\begin{align}
 n_T(\textbf{x},0)=\frac{N_T(\textbf{x})}{2}.
\end{align}

\subsection{Charge injection}
%%%%%%%%%%%%%%%%%%%%%%%%%%%%%%%%%%%%%%%%%%%%%%%%%%%%%%%%%%%%%
As has been discussed in our earlier work \cite{raftari2015self}, there are two 
possibilities within the DDR approach to model charge injection by a low- to 
moderate-energy electron beam via the source terms $S_n$ and $S_p$. 
The first fine-scale model captures the discrete nature of the electron beam.
The rate of particle production resolved at the level of pulses produced by 
individual PE impacts is given by:
\begin{align}
\label{equ:pulsedSource}
 S_{n,p}(\mathbf{x},t)= \sum_i\dfrac{g_{n,p}(\mathbf{x},E_{\rm lan})}{L(t_{\rm g})-L(t_{i})}\dfrac{dL}{dt}(t-t_{i}),
 \end{align}
where $L(t)$ is the logistic function, 
$i$ is the number of the particular individual PE, $t_{i}$ is the 
$i$-th PE impact time, and $t_{\rm g}$ is the 
generation time of the electron-hole pairs, whose choice is discussed 
in the next section.
With single-impact 
events, due to a relatively small number of produced pairs, the continuous results of the 
DDR model should be interpreted as probability densities rather than 
particle densities, especially at lower PE energies. 
 
The second model is designed for studying the sustained bombardment of the sample 
and is based on the temporal average of the above pulsed source function:
\begin{align}
\label{equ:continuousFocused}
 S_{n,p}(\mathbf{x},t)=\frac{j_0}{q}g_{n,p}(\mathbf{x},E_{\rm lan}),
\end{align}
where $j_0$ is the average electron beam current. 

Both source functions contain the semi-empirical distribution function of the 
charge pairs at the end of the initial generation stage:
\begin{align}\label{equ:Gus}
\begin{split}
g_{n,p}(\mathbf{x},E_{\rm lan}) =\left(A\frac{E_{\rm lan}}{E_{\rm i}}+B\right)
\frac{1}{\pi R^3}\exp\left(-\frac{7.5}{R^2}\vert\mathbf{x}-\mathbf{x}_0\vert^2\right)
 \end{split}
\end{align}
where $E_0$ and $E_{\rm lan}=E_0 + V_{\rm s}$ are the beam energy and the effective landing
energy of PE's, $V_s$ is the surface potential at the point of PE impact, 
$E_{\rm i}$ is the pair creation energy, 
$R$ is the maximum PE penetration depth (discussed further in detail),
$\mathbf{x}_0$ is the center of the Gaussian distribution with the distance 
of $0.3R$ from the sample-vacuum interface, and $A$ is the constant corresponding to 
the backscattering rate. In the hole distribution function $g_p$ the constant $B$ 
is zero, however, it is different from zero in the electron distribution function 
$g_n$ accounting for the remaining PE's. 

In the continuous irradiation mode we consider two additional modifications
of the source functions. One pertains to a defocused beam such that the 
computational domain is smaller than the beam radius.
In this case we use the following distribution function derived from (\ref{equ:Gus}) by integrating
over horizontal coordinates and enforcing the conservation of the amount of 
generated charge pairs:   
\begin{align}\label{equ:GusDefocused}
g_{n,p}(\mathbf{x}(r,z),E_{\rm lan}) =A_{n,p}^{'} \exp\left(-\beta\vert z-z_0\vert^2\right),
\end{align}
where
\begin{align}
 A_{n,p}^{'}=\frac{1-\exp(-\beta \delta^2)}{\beta \delta^2}A_{n,p},
\end{align}
\begin{align}
A_{n,p} =\left(A\frac{E_{\rm lan}}{E_{\rm i}}+B\right)\frac{1}{\pi R^3},~~~~ \beta=\frac{7.5}{R^2},
\end{align}
and $\delta$ is the radius of the irradiated area (computational domain) at the surface. 
Accordingly, the beam current can be calculated as
\begin{align}
\label{equ:currentDefocused}
j_0=i_0\pi\delta^2, 
\end{align}
where $i_0$ is the current density. The formula (\ref{equ:currentDefocused}) adjusts 
the beam current to achieve results independent of $\delta$.

If, on the other hand, the radius of a partially focused beam is smaller than 
the radius of the computational domain we resort to the following distribution:
\begin{align}
\begin{split}
&g_{n,p}(\mathbf{x}(r,z),E_{\rm lan}) =\frac{1}{\beta \delta^2+\exp(-\beta\delta^2)}A_{n,p} \times \\ 
&\begin{cases}
 \exp\left(-\beta\vert z-z_0\vert^2\right), & r\leq\delta\\
 \exp\left(-\beta(r^2+\vert z-z_0\vert^2)\right), & r>\delta.
\end{cases}
\end{split}
\end{align}
Here $\delta$ denotes the beam radius rather than the radius of the computational domain.

\subsection{Sample-vacuum interface and tertiary electrons}
%%%%%%%%%%%%%%%%%%%%%%%%%%%%%%%%%%%%%%%%%%%%%%%%%%%%%%%%%%%%%
The boundary conditions on $V$, $n$, $p$ at the interfaces 
of the sample with its holder and at the walls of the vacuum 
chamber are standard: Dirichlet at ohmic contacts and Neumann
to simulate isolation and prevent any currents from 
flowing through the corresponding interface.

The sample-vacuum interface, however, is not common in DDR-type simulations.
Previously \cite{raftari2015self} we have used a Robin-type boundary condition
$\textbf{J}_n\cdot\nu = v_e(n-n_i)$ for $n>n_{i}$ at this interface,
which sets the SE current density at the level proportional to 
the charge density at the boundary with the emission velocity
$v_{e}\leq v_{th}$ controlling the magnitude of the current ($\nu$ is the 
outward normal vector at the surface). 
As mentioned in the
Introduction this model does not account for the electrons that are
being pulled back to the sample by a positive surface potential --
the so-called tertiary electrons (TE's). This leads to nonphysical results
-- very strong positive charging of samples under prolonged irradiation
with low-energy beams.

Experiments show \cite{seiler1983secondary} that 
the energy of secondary electrons, although greater than the electron affinity of the material, 
rarely exceeds $10$~eV. Therefore, even a relatively weak positive potential 
at the surface will pull back some of the emitted secondary electrons. 
To account for this tertiary electron current we propose the 
following modified version of the Robin-type boundary condition
at the sample-vacuum interface:
\begin{align}
\label{equ:interface1}
\textbf{J}_n\cdot\nu&=\begin{cases}
                       v_e(n-n_i)-\alpha \dfrac{\partial V}{\partial \nu}^-, & \text{if}\ \ \ n> n_i;\\
                       0,  & \text{otherwise},
                      \end{cases}\\
\label{equ:interface2}
\textbf{J}_p\cdot\nu&=0, \ \ \text{on}\ \ \Sigma_2\times[0,t_{\rm end}],
\end{align}
where 
\begin{align}
 \dfrac{\partial V}{\partial \nu}^-|_{\Sigma_2}=\begin{cases}
                       \dfrac{\partial V}{\partial \nu}|_{\Sigma_2}, & \text{if}\ \ \ \dfrac{\partial V}{\partial \nu}<0;\\
                       0,  & \text{otherwise},
                      \end{cases}
\end{align}
\begin{align}
\label{equ:alphaFun1}
 &\alpha(\text{max}(V^+))=\\ \nonumber
 &\begin{cases}
                       0, & \text{if}\ \ \ \text{max}(V^+)<V_{min};\\
                       \alpha_{max}\frac{\text{max}(V^+)-V_{min}}{V_{max}-V_{min}}, & \text{if}\ \ \ V_{min}\leq\text{max}(V^+)<V_{max};\\
                       \alpha_{max},  & \text{otherwise},
                      \end{cases}
\end{align}
\begin{align}
 V^+|_{\Sigma_2}=\begin{cases}
                       V|_{\Sigma_2}, & \text{if}\ \ \ V>0;\\
                       0,  & \text{otherwise},
                      \end{cases}
\end{align}
\begin{align}
 \text{max}(V^+) = \text{Maximum of}~~(V^+|_{\Sigma_2}-V_g),
\end{align}
and
\begin{align}
\label{equ:alphaFun2}
 \alpha_{max}=\frac{v_e\displaystyle\int_{\Sigma_2}(n-n_i)dA}{\displaystyle\int_{\Sigma_2}\frac{\partial V}{\partial \nu}dA},
\end{align}
where $V_g$ is the applied potential at the upper boundary, which in the 
present study is set to zero ($V_g=V|_{\Sigma_1}=0~V)$. 
The term $-\alpha \frac{\partial V}{\partial \nu}^-$ in (\ref{equ:interface1}) 
represents the tertiary electrons current density. 
The function $\alpha$ controls the total magnitude of this current and the factor 
$-\frac{\partial V}{\partial \nu}^-$ determines its spatial distribution.
We assume that tertiary electrons will re-enter the sample only through regions 
where the normal component of the electric field is negative. The stronger is the
local attractive electric field, the higher is the density of tertiary current at that location.
 
The function $\alpha(t)$ is chosen here in such a way, see (\ref{equ:alphaFun1})--(\ref{equ:alphaFun2}), 
that the magnitude of the total tertiary current varies linearly from zero, 
when the maximum surface potential $V^{+}(t)$ is below a certain value $V_{min}$,
to the value of the total outward SE current, when $V^{+}(t)$ reaches $V_{max}$.
This means that the net current through the sample-vacuum interface will be zero
if  $V^{+}(t)\geq V_{max}$ as all SE's leaving the sample will re-enter the sample 
as tertiary electrons. Typically this leads to the surface potential never raising 
above $V_{max}$ (or $V_{max}+V_{g}$). 
This choice of $\alpha(t)$ is not unique and could be further refined to take the 
energy spectrum of the SE's into account.
One should also mention that, from the computational point of view, 
the mesh along the sample-vacuum interface should be fine enough in order to capture 
the gradient of the potential at the surface.

\subsection{Numerical solution}
%%%%%%%%%%%%%%%%%%%%%%%%%%%%%%%%%%%%%%%%%%%%%%%%%%%%%%%%%%%%%
The first step toward obtaining a numerical solution of an equation or a system of 
equations is to investigate the existence and uniqueness of the solution. With regards to 
the present model, the consistency analysis relies on previously published results. 
A detailed investigation concerning the existence and uniqueness of stationary 
drift-diffusion equations can be found in \cite{markowichsemiconductor}. In a study 
conducted by Jerome \cite{jerome1985consistency} a mathematical analysis of a system 
solution map for the weak form of the DDR model, which forms a basis for the numerical 
solution of the model, has been provided. Also, in a follow-up study 
by Busenberg et al. \cite{busenberg1993modeling} the wellposedness of a DDR model 
similar to the present one (with different source/sink terms) has been demonstrated. 

The multiscale nature of the problem calls for the same strategy as was used 
in \cite{raftari2015self} regarding the scaling of variables. We apply the finite element 
method (FEM) for the numerical solution of the model equations and implement it as a 
solver within the COMSOL Multiphysics package. To balance the accuracy and the 
computational costs a careful strategy is needed. Our investigations show that the 
best (i.e., most reliable) results are obtained when we use adaptive (or local) mesh refinement, 
Lagrange shape functions, the fully coupled approach with the 
Newton-Raphson solver, and an adaptive time-stepping algorithm. 
The use of the adaptive grid refinement, although costly, alleviates the need
for more sophisticated approaches, such as the traditional exponential fitting applied in 
semiconductor studies \cite{polak1987semiconductor,ten1993exponential}. 
To reduce the computational cost of the adaptive mesh refinement, 
a simple strategy has been followed by using 
a combination of both adaptive and local refinement methods. 
Namely, the adaptive refinement was applied only in the initial simulations to 
identify the regions where a fine mesh is needed and 
then the local refinement is used in the follow-up simulations. 

In some cases, we reduce the 
original 3D problem to a 2D problem in the $(r,z)$-plane of the cylindrical coordinate 
system as the geometry, boundary conditions, and the source are all axially symmetric.
In the simulations of beam scanning over laterally inhomogeneous samples a fully three-dimensional
implementation was used.

%%%%%%%%%%%%%%%%%%%%%%%%%%%%%%%%%%%%%%%%%%%%%%%%%%%%%%%%%%%%%
\section{\label{sec:calibration}Calibration and comparison}
%%%%%%%%%%%%%%%%%%%%%%%%%%%%%%%%%%%%%%%%%%%%%%%%%%%%%%%%%%%%%
In this section we 
provide a detailed account of the calibration 
procedure for alumina and silica, 
which employs experimental data from the studies by Dawson \cite{dawson1966secondary} 
and Young et al\cite{yong1998determination},
simultaneously explaining the underlying approximations of the DDR approach.
We also compare our time-domain simulations with the results of an
alternative one-dimensional approach.

\subsection{Reproducing the standard yield of insulators}
There are two main kinds of yield measurements from insulators: dedicated measurements
with homogeneous pure samples \cite{dawson1966secondary, yong1998determination} and SEM scans of 
insulator-containing targets \cite{kim2010charging}.
In the former case often a great care is taken to avoid the charging effects. Typically,
a defocused beam, a weak beam current, and a pulse of short duration are used. 
We define the SE yield free of charging effects as the {\it standard yield}
and calibrate our code to reproduce such data as close as possible.

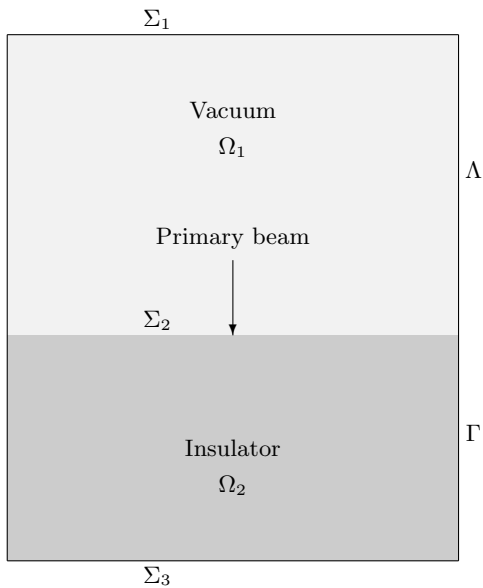
\begin{figure}[t!]
\begin{tikzpicture}
% \hspace{+cm}
     \fill[gray!10!white] (-3,0) rectangle (3,7);
    % \fill[gray!70!white] (2.5,1.5) arc (0:180:2.5cm);
     \fill[gray!40!white] (-3,0) rectangle (3,3);
%      \draw (2,2) arc (0:180:2cm);
     \setlength{\unitlength}{1cm}
     \put(0,4){\vector(0,-1.5){1}}
     \node[] at (0,4.3) {Primary beam};
     \node[] at (0,5.5) {$\Omega_1$};
     \node[] at (0,6) {Vacuum};
     \node[] at (0,1) {$\Omega_2$};
     \node[] at (0,1.5) {Insulator};
     \node[] at (-1,7.2) {$\Sigma_1$};
     \node[] at (-1,3.2) {$\Sigma_2$};
     \node[] at (-1,-0.2) {$\Sigma_3$};
     \node[] at (3.2,5.2) {$\Lambda$};
     \node[] at (3.2,1.7) {$\Gamma$};
%          \node[] at (0,2.6) {$\Omega_2$};
%     \node[] at (0,3) {\scriptsize{Insulator}};
%      \node[] at (3.4,2.2) {$\partial\Omega_2$};
%      \node[] at (-3.4,2.2) {$\partial\Omega_2$};
%      \node[] at (0,-0.3) {$\partial\Omega_2$};
%      \node[] at (-3.4,4.5) {$\partial\Omega_1$};
%      \node[] at (3.4,4.5) {$\partial\Omega_1$};
%      \node[] at (0,5.2) {$\partial\Omega_1$};
     \draw(-3,0)  -- (-3,7);
     \draw(3,0)   -- (3,7);
     \draw(-3,0)  -- (3,0);
     \draw(-3,7)   -- (3,7);
%      \draw [black](-3,4.5) -- (3,4.5);
%      \draw (0,4.5)--(0,3.7);
%       \draw [decorate,decoration={brace,amplitude=4pt},xshift=0.1cm,yshift=0pt]
%       (0,4.5) -- (0,3.7) node [midway,right,xshift=.1cm] {$0.3R$};
%       \fill[black] (0,3.7) circle (0.05);
\end{tikzpicture}
\caption{The schematic representation of the model.}
\label{fig2}
\end{figure}

Parameters of standard-yield experiments\cite{dawson1966secondary, yong1998determination}
(current, pulse duration, irradiation area) imply that the probability for the primary 
electrons to land anywhere close to each other on the surface is very small. 
In fact, with defocused beams and low, short-duration currents the expected 
distance between PE's is large enough to permit neglecting mutual interaction 
between any two impact zones. This is the main reason why standard-yield 
measurement are free of charging effects.
The single-impact source function (\ref{equ:pulsedSource}) with $i=1$
allows to compute directly the expected number of emitted SE's per single 
(isolated) PE impact and is, therefore, applicable for modeling the standard yield. 
The cross-section of the axially-symmetric configuration 
used for calibrating the code is shown in Fig.~\ref{fig2}. With sufficiently 
large computational domain the boundary conditions at the sides of the sample have
no influence on the SE yield from a single PE impact and were set to
Neumann (zero current).
%%%%%%%%%%%%%%%%%%%%%%%%%%%%%%%%%%%%%%%%%%%%%%%%%%%%%%%%%%%%%%%%%%%%%%%%%%%%%%%%%%%%%%%
%%%%%%%%%%%%%%%%%%%%%%%%%%%%%%%%%%%%%%%%%%%%%%%%%%%%%%%%%%%%%%%%%%%%%%%%%%%%%%%%%%%%%%%

There are two classes of parameters that may be tuned within their 
physically admissible ranges: those that determine the shape of the source 
function approximating the initial pair generation and the short-time high-energy transport stage, and the
material (bulk) parameters that determine the transport and trapping/de-trapping
at much longer time scales. While these time scales may seem well-separated,
in the DDR model material parameters, especially the emission velocity $v_{e}$, 
have some influence on the initial transport stage as well.

\begin{figure}[t!]
\includegraphics[scale=0.62] {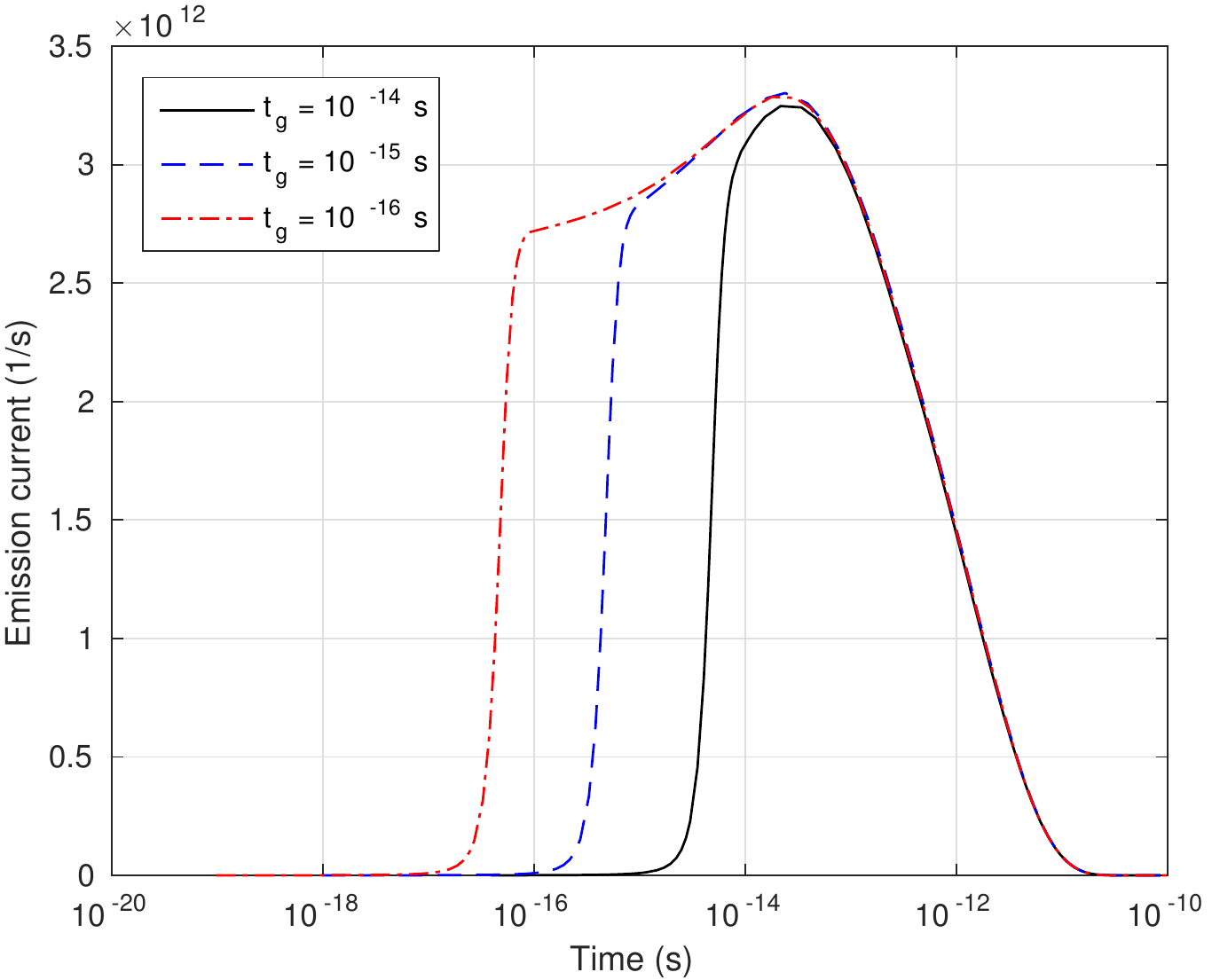}
\caption{Calibrated emission current pulse from sapphire sample after the impact of
a single $500$~eV PE. Corresponding standard yields are: 
$6.07$ ($t_{\rm g}=10^{-14}$~s), $6.17$ ($t_{\rm g}=10^{-15}$~s), and 
$6.14$ ($t_{\rm g}=10^{-16}$~s).}   
\label{fig:currentSingle}
\end{figure}

The pair generation time $t_{\rm g}$, defined as the time when all pairs have 
already been generated, determines the time width of the pulsed source 
functions $S_{n,p}({\mathbf x},t)$ and of the resulting SE emission current pulse. 
According to theoretical and experimental investigations
by D.I.~Vaisburd et al \cite{vaisburd2008poole}, between $10^{-17}$ and $10^{-14}$~s 
after impact the generated secondary pairs have already lost their ability to ionize the medium
and their energy spectrum begins to evolve away from the spectrum 
of the primary beam as the result of collisions. However, up to $10^{-14}$~s 
most of the generated pairs still have energies above 20~eV. 
Since ``true'' SE's dominating the emission spectrum have energies below $20$~eV,
most of them must be emitted after $10^{-14}$~s. It has also been found that $10^{-11}$~s 
after impact all generated pairs are already thermalized with their energy 
spectra tightly localized around the edges of conduction and valence bands and 
trapping becomes more pronounced. 
Hence, the SE emission current pulse following 
a single PE impact should start after $10^{-17}$~s and be almost 
finished by $10^{-11}$~s. Moreover, if one aims at modeling ``true'' SE's, then 
the relative contribution to the total 
emission between $10^{-17}$ and $10^{-14}$~s should be small, compared
to the contribution between $10^{-14}$ and $10^{-11}$~s.
The DDR method produces exactly this type of pulses
for $t_{\rm g}$ set between $10^{-16}$ and $10^{-14}$~s, 
see Fig.~\ref{fig:currentSingle}. Notably, other parameters influence only 
the magnitude, not the duration of the emission current pulse.

We emphasize that the curves of 
Fig.~\ref{fig:currentSingle} should be interpreted in the probabilistic sense.
Namely, the integral of this curve between any two time instants 
$t_{A}\leq t_{B}$ is the number of particles expected to be emitted from the 
sample surface during the corresponding time interval.
Thus, the expected yield at a given PE energy can be computed by numerically 
integrating the emission current between $t=0$ and some sufficiently large $t> 10^{-11}$.
%%%%%%%%%%%%%%%%%%%%%%%%%%%%%%%%%%%%%%%%%%%%%%%%%%%%%%%%%%%%%%%%%%%%%%%%%%%%%%%%%%%
%%%%%%%%%%%%%%%%%%%%%%%%%%%%%%%%%%%%%%%%%%%%%%%%%%%%%%%%%%%%%%%%%%%%%%%%%%%%%%%%%%%

Among the material parameters the carrier mobilities $\mu_{n,p}$ have been 
determined with the highest precision and are simply assumed here to have 
the same values as in \cite{raftari2015self,hughes1979generation,hughes1975hot,renoud2004secondary}. 
Strictly speaking, these are the so-called low-energy mobilities and a more 
rigorous approach would be to use femtosecond and picosecond mobilities 
to model the transport of particles during the corresponding time intervals 
after the impact \cite{vaisburd2008poole}.
However, mainly due to the absence of data about these high-energy mobilities, 
here we use the same low-energy mobility values at all times. 
Nevertheless, the extremely short duration $[10^{-17},10^{-12}]$~s of 
the high-energy regime allows us to expect the approximations made in the 
DDR approach concerning the mobility values during this stage
to be appropriate at least in the numerical sense. Notice that the changes 
in the yield do not exceed $0.1$ when we vary the generation time $t_{\rm g}$ 
between $10^{-16}$ and $10^{-14}$~s in Fig.~\ref{fig:currentSingle}. 
Thus, to have any significant impact on the yield the mobility would 
have to vary dramatically during this interval of time.

Parameters $\sigma_{n,p}$ and $N_{T}$ related to trapping weakly influence the 
magnitude of the emission current pulse and have, generally,
large uncertainties. For example, in a study set out to investigate electron 
trapping in alumina \cite{pickard1970analysis} 
a relatively large variation of $10^{-21}$ to $10^{-15}$ $\text{cm}^2$ was reported for 
the electron capture cross section $\sigma_{n}$ in polycrystalline alumina. 
The same study also revealed that polycrystalline 
metal oxide materials like sapphire ($\alpha$-alumina) generally have trap 
site densities $N_{T}$ in the order of $10^{18}~\text{cm}^{-3}$. Insulating solids are often 
grouped into three types: crystals, polycrystalline 
and amorphous\cite{cazaux1996electron}. The trap site density has been estimated 
to be around $10^{16}~\text{cm}^{-3}$ for an alumina crystal, from $10^{17}$ to 
$10^{20}~\text{cm}^{-3}$ for polycrystals, and around $10^{21}~\text{cm}^{-3}$ 
for an amorphous sample. 

Probably one of the most comprehensive and systematic studies on charge transport 
and trapping in silica has been done by DiMaria and co-workers \cite{dimaria1979radiation,dimaria1989trap,buchanan1989coulombic,dimaria1991trapping}, 
where a strong link has been identified between the capture cross sections and the 
nature of traps and the capture cross sections have been estimated to 
range from $10^{-18}$ to $10^{-13}$ $\text{cm}^2$. 
Confusingly, the values and ranges for these parameters are not limited 
to the above mentioned estimates \cite{ning1976capture,ning1976high,williams1965photoemission}. 

Another parameter that strongly influences the magnitude of the emission current pulse 
is the maximum PE penetration depth $R$. It determines the spatial shape of 
the source functions $S_{n,p}({\mathbf x},t)$ and, therefore, the expected number of particles
in the neighborhood of the sample-vacuum interface -- the main contributors to the 
emission current.
Many semi-empirical expressions have been proposed for $R$ with the following 
general form $R(\rho,E_0)=CE_0^{\Gamma}$,
where the values for $C$ and $\Gamma$ vary from study to study\cite{fitting2011secondary,fitting1977electron}. 
The constant $C$ depends on the material and the exponent $\Gamma$ has been mostly assumed to have 
a certain material-independent value, although, in some studies $\Gamma$ 
has also been considered material-dependent\cite{feldman1960range}.

\begin{figure*}[t!]
\includegraphics[scale=0.62] {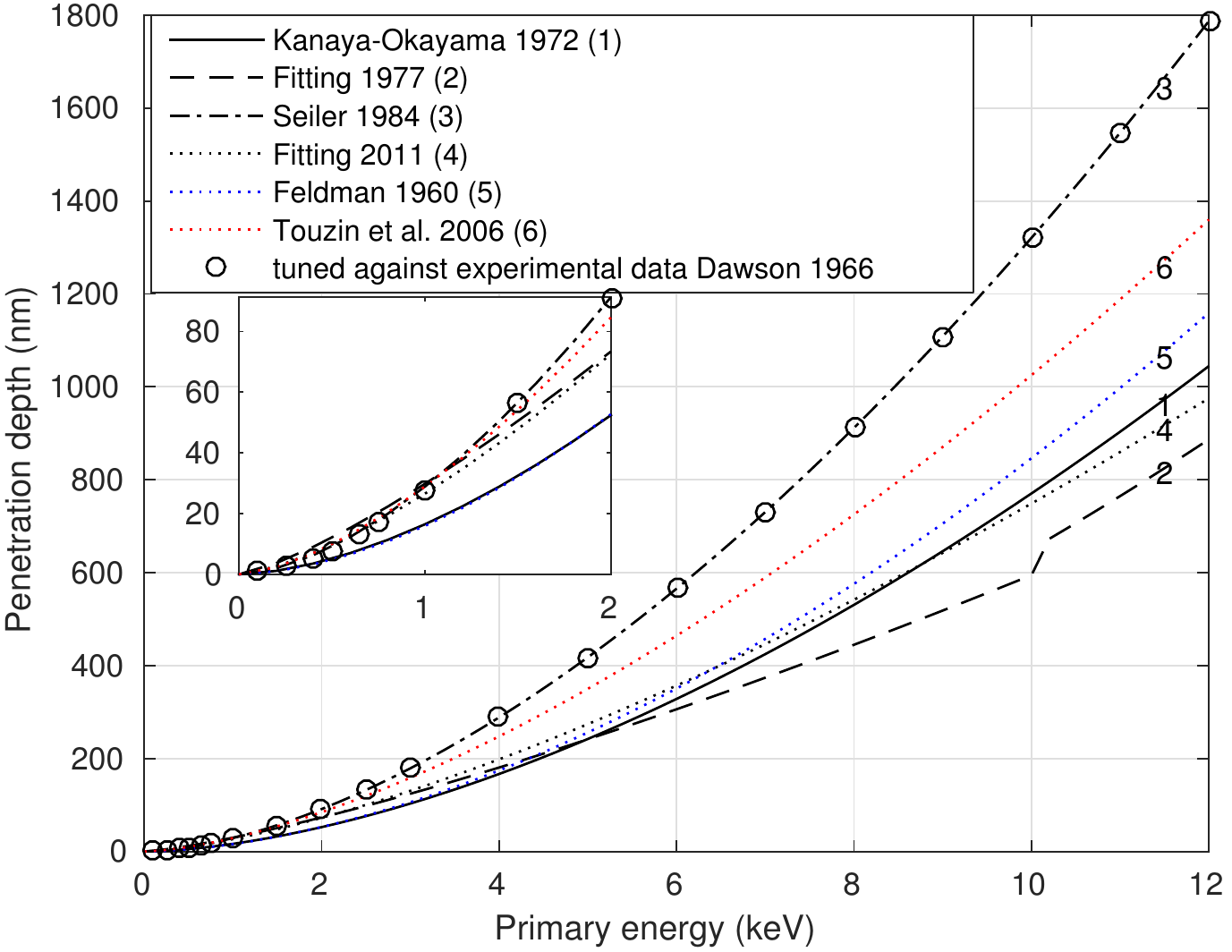}
\includegraphics[scale=0.62] {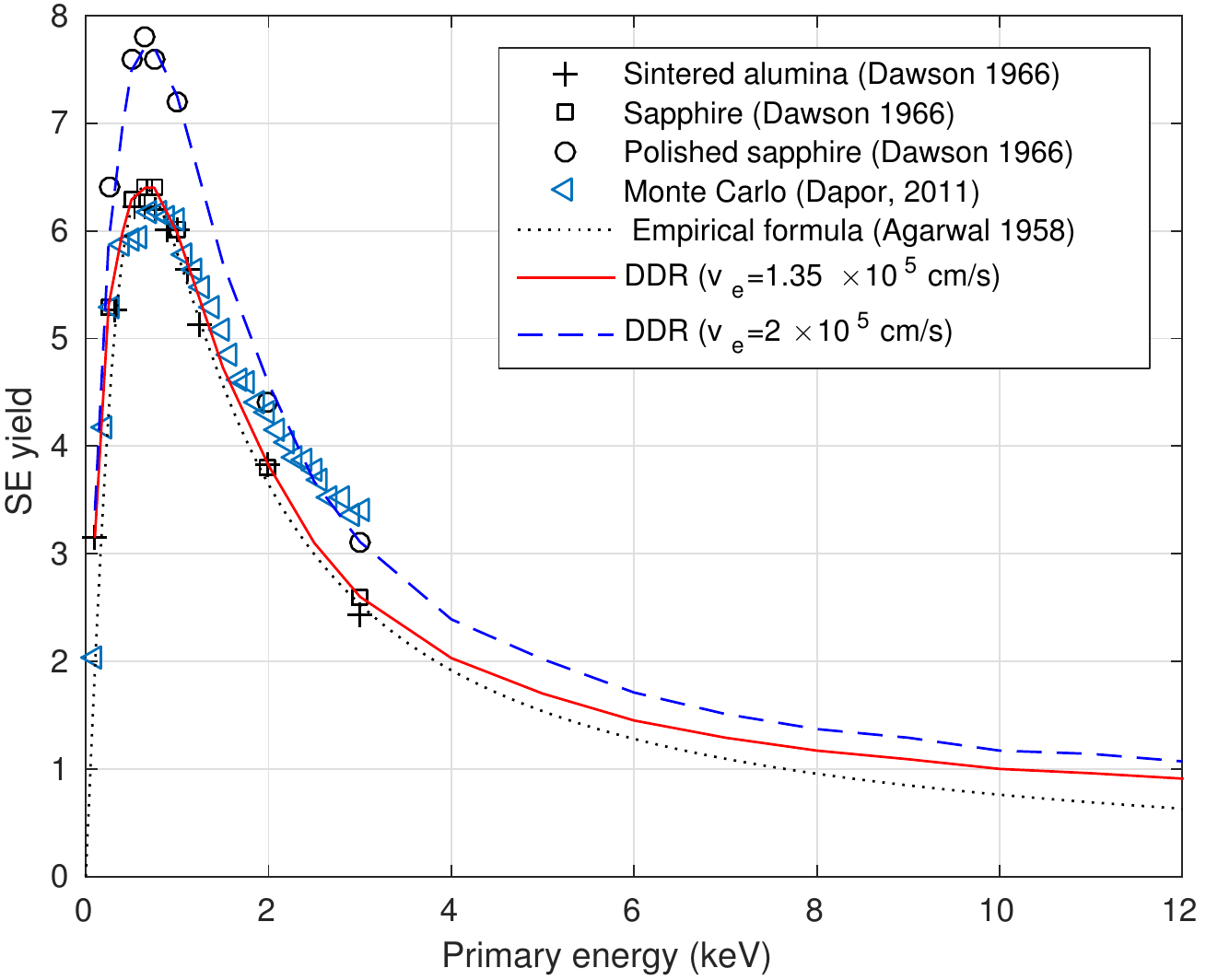}
\caption{Sapphire: penetration depth (left) and 
SE yield (right) as functions of PE energy.}   
\label{fig:sapphireSingle}
\end{figure*}

\begin{figure*}[t!]
\includegraphics[scale=0.62] {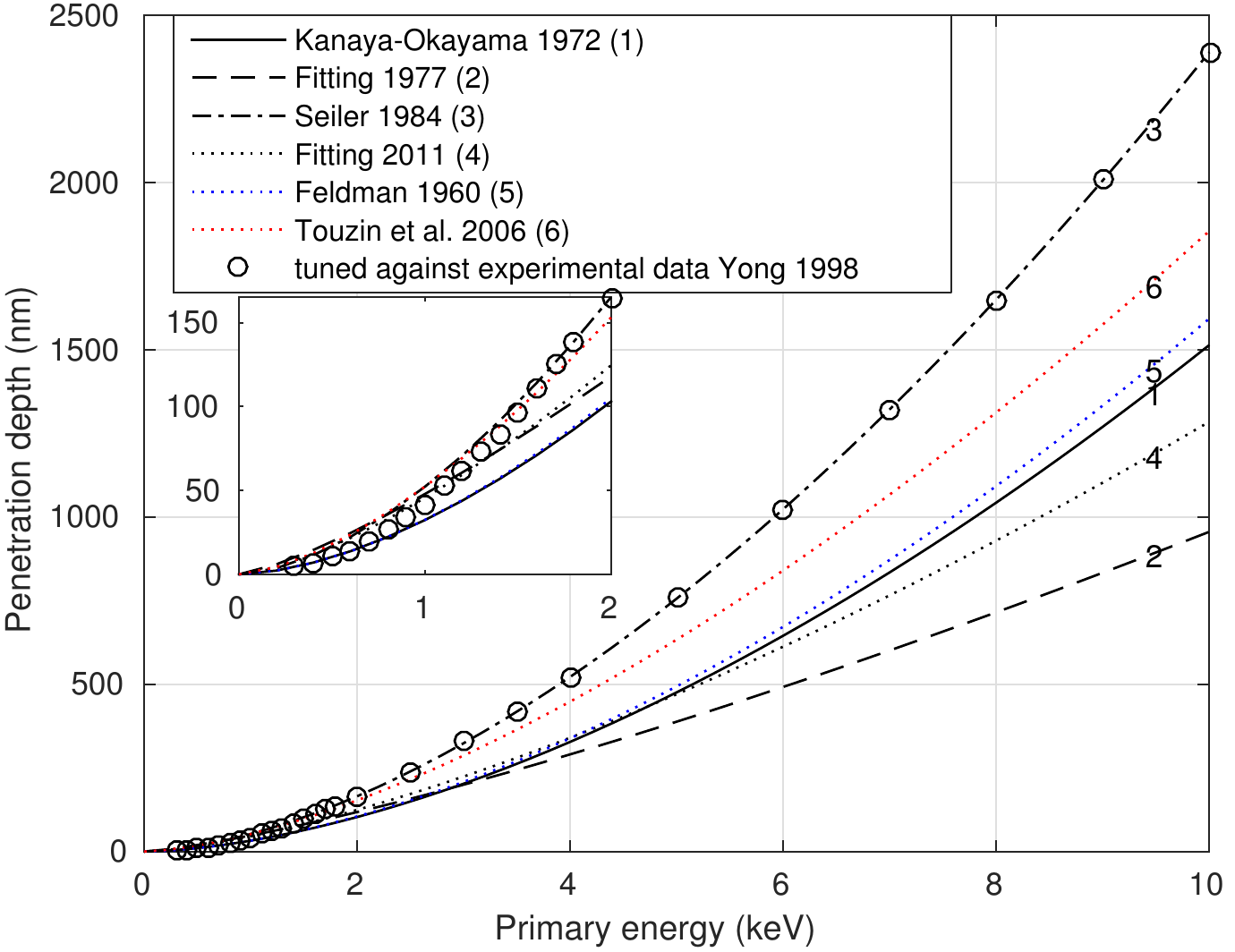}
\includegraphics[scale=0.62] {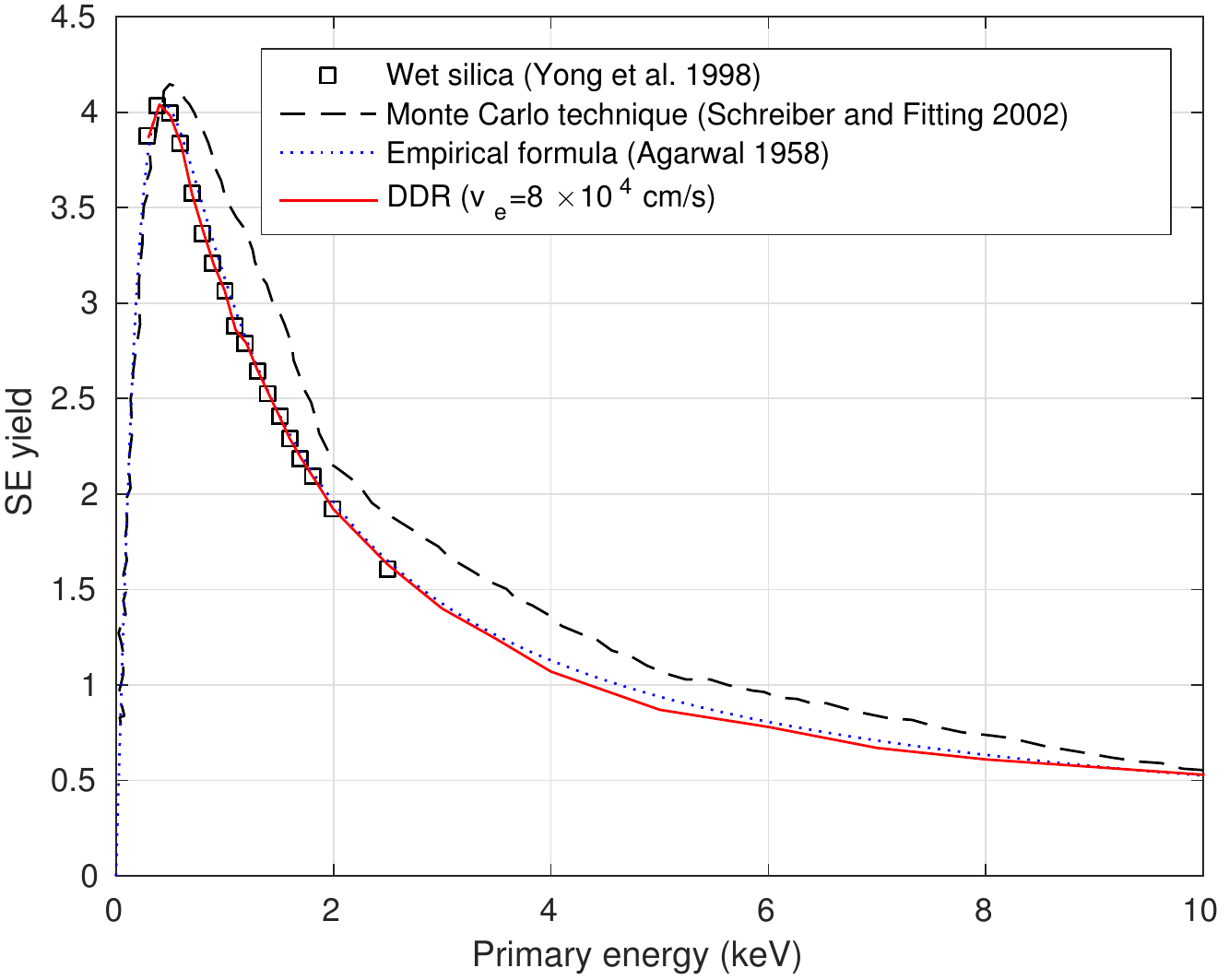}
\caption{Silica: penetration depth (left) and 
SE yield (right) as functions of PE energy.}   
\label{fig:silicaSingle}
\end{figure*}

The exponential expression for $R$ emanates from Bethe's theory 
for the stopping power of charged particles in matter. 
Bethe's formula involves the density, atomic number, and atomic weight of the material. 
However, with the exception of studies by Kanaya and Okayama \cite{kanaya1972penetration} and 
by Feldman \cite{feldman1960range}, the density of the 
material is considered to be the only parameter influencing the electron penetration depth. 
As of now the estimation of $R$ is far from being certain as can be seen from large discrepancies 
in the penetration depth estimates employed by different authors, see Fig.~\ref{fig:sapphireSingle}~(left) 
and Fig.~\ref{fig:silicaSingle}~(left). Apparently, similar disagreement concerning the penetration depth 
exists for metals as well\cite{cosslett1964multiple}.

%%%%%%%%%%%%%%%%%%%%%%%%%%%%%%%%%%%%%%%%%%%%%%%%%%%%%%%%%%%%%%%%%%%%%%%%%%%%%%%%%%%
%%%%%%%%%%%%%%%%%%%%%%%%%%%%%%%%%%%%%%%%%%%%%%%%%%%%%%%%%%%%%%%%%%%%%%%%%%%%%%%%%%%
%%%%%%%%%%%%%%%%%%%%%%%%%%%%%%%%%%%%%%%%%%%%%%%%%%%%%%%%%%%%%%%%%%%%%%%%%%%%%%%%%%%
%%%%%%%%%%%%%%%%%%%%%%%%%%%%%%%%%%%%%%%%%%%%%%%%%%%%%%%%%%%%%%%%%%%%%%%%%%%%%%%%%%%

Having identified $v_{e}$, $\sigma_{n,p}$, $N_T$, and $R$ as the most uncertain 
of the model parameters influencing the magnitude of the emission current pulse 
we have performed a series of numerical experiments
to determine the sensitivity of the DDR model output (SE yield) 
with respect to changes in these parameters.
During these simulations some of the factors would be held
fixed while other were varied with the goal to achieve the best 
possible fit between the computed SE yields and the experimental data.       
Three key points emerged from this analysis:
\begin{itemize}
 \item The shape of the yield-energy curve is influenced 
by the capture cross section and the density of traps. Namely, 
the larger are the trap density and the capture cross section, 
the lower is the high-energy tail of the curve.
\item The emission velocity affects the height not the shape of the yield-energy curve.
\item Following any one of the published penetration depth formulas together with 
adjusting the values of material parameters within their permitted ranges 
does not produce yield-energy curves fully compatible with the 
experimental data over the whole range of PE energies.
\end{itemize} 

In view of these facts and the aforementioned uncertainty about 
the energy dependence of the penetration depth, fine-tuning $R$ for each 
PE energy against the available experimental data was deemed by us as not only 
admissible, but also necessary. 
While tuning $R$ other parameters have been fixed at the best found 
fitting values within their reported ranges. 
In particular, the electron and hole capture cross-sections were set at 
the frequently used values of $10^{-15}~\text{cm}^2$ and $10^{-17}~\text{cm}^2$, 
respectively. The trap site density turned out to be slightly higher than the 
reported upper bound $10^{19}~\text{cm}^{-3}$, namely, 
$3\times10^{19}~\text{cm}^{-3}$, leading to the initial 
(equilibrium) density of trapped electrons of $1.5\times10^{19}~\text{cm}^{-3}$, 
close to what was used by us previously\cite{raftari2015self}. 

For PE energies higher than $2$ keV the tuned penetration depths 
for sapphire and silica 
presented in Fig.~\ref{fig:sapphireSingle}~(left) and 
Fig.~\ref{fig:silicaSingle}~(left), perfectly 
match those of Lane and Zaffarano \cite{lane1954transmission} and are
well-described by the formula of Young \cite{young1956penetration}: 
\begin{align}
\label{equ:depth2}
R(\rho,E_0)=115\frac{E_0^{1.66}}{\rho}~[\text{nm}],~~~~~E_0\geq 2~\text{keV}.
\end{align}
However, according to Young \cite{young1956penetration} the exponent of $E_{0}$ is $1.35$, 
while the present results agree with the earlier reported \cite{lane1954transmission} value 
of $1.66$. There is some argument about this exponent in the literature. 
For instance, the study about Kapton and Teflon \cite{yang1987electron} supports the idea of $1.66$. 
Yet, the investigation of Salehi and Flinn \cite{salehi1980experimental} with 
$\text{V}_2\text{O}_5-\text{P}_2\text{O}_5$ materials shows that, although 
at low energies the exponent is close to $1.35$, neither $1.35$ nor $1.66$ 
provide good matches with higher-energy experimental data. 
The value of $1.66$ was assumed for sapphire in several other 
investigations as well\cite{melchinger1995dynamic,belhaj2000time}.  

As can be seen from the insets of Fig.~\ref{fig:sapphireSingle}~(left) and 
Fig.~\ref{fig:silicaSingle}~(left) at energies below $2$~keV the tuned penetration 
depths deviated from the formula (\ref{equ:depth2}) and did not follow any other published formulas, 
while remaining within their range. Least squares fitting of a separate exponential 
formula of the type (\ref{equ:depth2}) to the tuned penetration depths 
for alumina and silica did not provide a satisfactory fit. This suggests that
below $2$~keV the energy exponent $\Gamma$ is indeed material dependent.
Hence, for calibration purposes penetration depths bellow $2$~keV must 
be determined by fitting to the corresponding standard yield data,
whereas above this energy the depth may be safely deduced from 
the formula (\ref{equ:depth2}).

With the tuned penetration depths the DDR method provides practically exact 
yield-energy curves for the whole range of PE energies. 
As was mentioned previously, the height of the yield-energy curve
is mainly controlled by the electron emission velocity at the vacuum-sample interface.
In Fig.~\ref{fig:sapphireSingle}~(right) we compare the output of the calibrated DDR model
with the standard-yield data \cite{dawson1966secondary} (reported also in the database of Joy \cite{Database}), 
as well as Monte-Carlo simulations \cite{dapor2011secondary} and the empirical formula of Agarwal \cite{agarwal1958variation} for alumina samples.
As far as the DDR model is concerned the only difference between the unpolished and polished 
alumina samples is the electron emission velocity at the sample-vacuum interface
($1.35\times10^5$~cm/s and $2\times10^5$~cm/s, respectively), which sounds reasonable, 
since surface polishing should not affect the maximum penetration depth. 

Comparison of the results by the calibrated DDR model with the experimental data \cite{yong1998determination}, 
Monte-Carlo simulations \cite{schreiber2002monte}, and the formula of Agarwal \cite{agarwal1958variation} 
for a silica sample is shown in Fig.~\ref{fig:silicaSingle} (right). The tuned electron emission velocity
at the silica-vacuum interface ($0.8\times10^5$~cm/s) is lower than the emission velocity 
at the alumina-vacuum interface, indicating that $v_{e}$ depends on both the material 
and the surface properties. 

DDR simulations indicate that the first and the second unit 
yields for sapphire occur around $50$~eV and $10$~keV, respectively. For silica, 
the unit yields are observed below $50$~eV and again at $4.35$~keV. 
The values of the calibrated material parameters used in 
the present study are listed in Table \ref{tab:I}.

%**********************************************************************************
\begin{table}[t]
\centering
\caption{Parameters of dielectric materials.}
    \begin{tabular}{|l|l|l|l|}
   \hline
    \small{Parameter}& $\text{SiO}_2$ & $\text{Al}_2\text{O}_3$& Unit\\
    \hline
    &&&\\
    $\varepsilon$ & $3.9$  & $10$  & \\
    $\mu_n$ & $20$ & $4$ & $\text{cm}^2\text{V}^{-1}\text{s}^{-1}$  \\
    $\mu_p$ & $0.01$ & $0.002$   &$\text{cm}^2\text{V}^{-1}\text{s}^{-1}$  \\
    $\sigma_n$& $10^{-15}$ & $10^{-15}$  & cm$^2$ \\
    $\sigma_p$ & $10^{-17}$ & $10^{-17}$ & cm$^2$  \\
    $v_{th}$ & $10^7$ &  $10^7$ &cms$^{-1}$  \\
    $\rho$ & $2.2$ & $3.98$ &  gcm$^{-3}$  \\
    $E_i$ & $28$ & $28$ &eV  \\
    \hline
    $N_t$  & $3\times10^{19}$ &  $\begin{array}{ll}
                                   3\times10^{19} & \text{sapphire}\\
                                   10^{20}  & \text{amorphous}
                                  \end{array}$ & cm$^{-3}$\\
    \hline                                  
    $v_e$  & $0.8\times10^5$ &  $\begin{array}{ll}
                                   1.35\times10^5 & \text{unpolished}\\
                                   2.0\times10^5    & \text{polished}\\
                                   1.4\times10^5  & \text{amorphous}
                                  \end{array}$
    & cms$^{-1}$\\
    \hline
    \end{tabular}
\label{tab:I}
\end{table}
%*******************************************************************************************

%%%%%%%%%%%%%%%%%%%%%%%%%%%%%%%%%%%%%%%%%%%%%%%%%%%%%%%%%%%%%
\subsection{\label{sec:defocused}Continuous irradiation with defocused beams}
%%%%%%%%%%%%%%%%%%%%%%%%%%%%%%%%%%%%%%%%%%%%%%%%%%%%%%%%%%%%%
Sustained bombardment, even with defocused beams, increases the probability
for an incoming PE to fall in a close proximity to a previous impact zone.
This will introduce the interaction between the previously trapped charges
and the newly generated pairs, so that the yield will vary with time.

At the moment a standard experimental procedure for measuring yield variation during 
sustained bombardment does not exist. Therefore, here we compare
predictions of the DDR model with the earlier one-dimensional simulations 
by the Flight-Drift (FD) model -- 
a self-consistent approach by Touzin et al\cite{touzin2006electron}. 
FD model is a current-density based formalism incorporating a detailed recombination 
and trapping mechanism.
For comparison purposes we have considered the same material (amorphous alumina), 
current density, and the penetration depth formula (energy exponent in (\ref{equ:depth2}) 
is set as $\Gamma=1.55$). We switch now to the continuous (time-integrated) 
source function (\ref{equ:continuousFocused}), 
(\ref{equ:GusDefocused})--(\ref{equ:currentDefocused}) suitable for long-time 
modeling.
 
Since the sample is amorphous alumina rather than sapphire, 
we choose a higher trap density of $10^{20}~\text{cm}^{-3}$ pertaining 
to the so-called shallow traps\cite{touzin2006electron}. We set the emission 
velocity to $1.4\times10^5$~cm/s, close to what we have obtained above for unpolished sapphire. 
We note that in time-domain investigations the quantity of interest is not the
charge yield, but the instantaneous ratio of the net SE emission current to the 
incident beam current -- SE emission rate.

\begin{figure}[t!]
\centering
\includegraphics[scale=0.62] {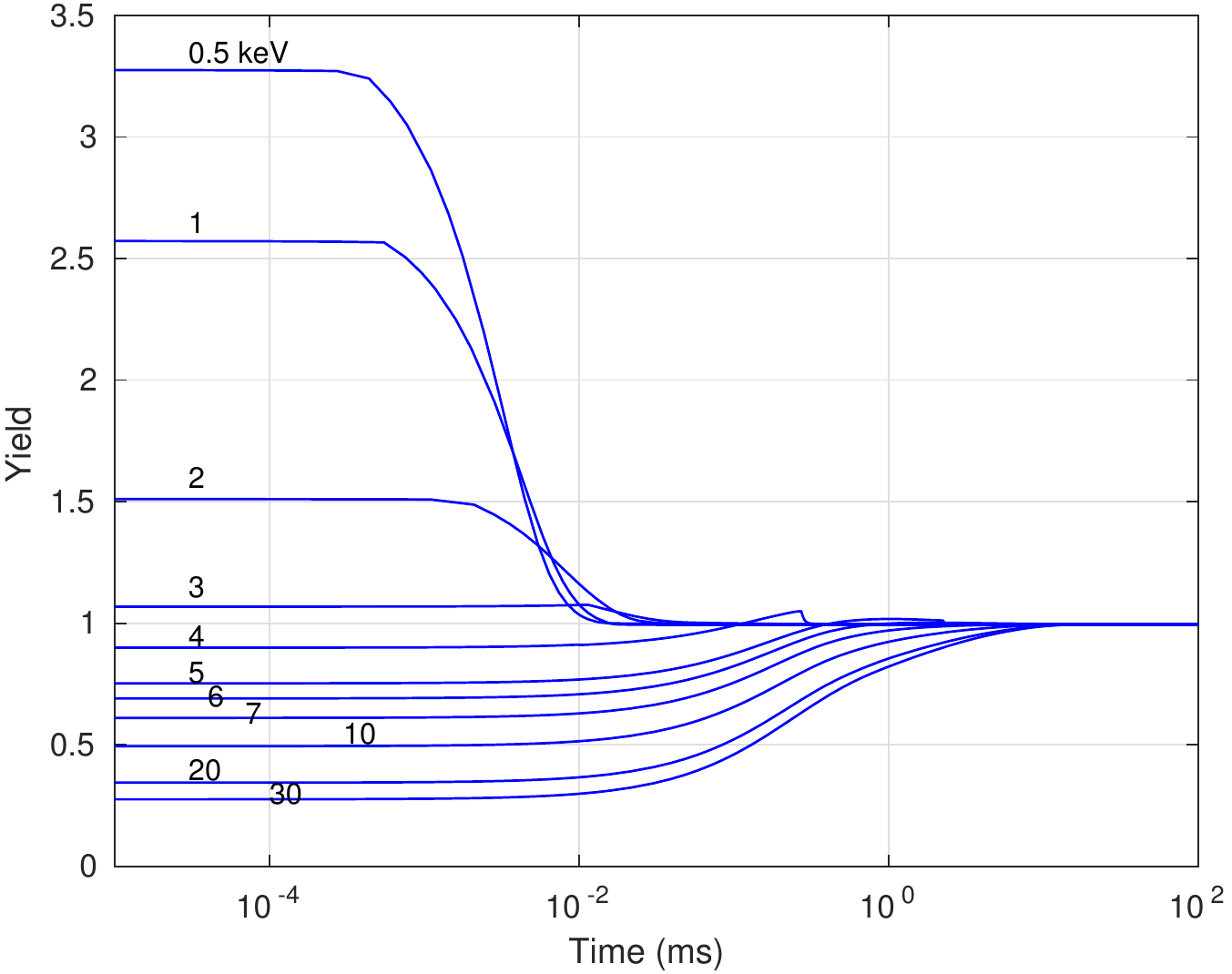}
\caption{Time evolution of the yield from an amorphous
alumina sample versus PE energy for a defocused beam with current density of $10^{-5}$~A/cm$^2$. 
The distance between $\Sigma_1$ and $\Sigma_2$ is $1$~mm.}   
\label{fig:amorAluminaYield}
\end{figure}

\begin{figure}[t!]
\includegraphics[scale=0.62] {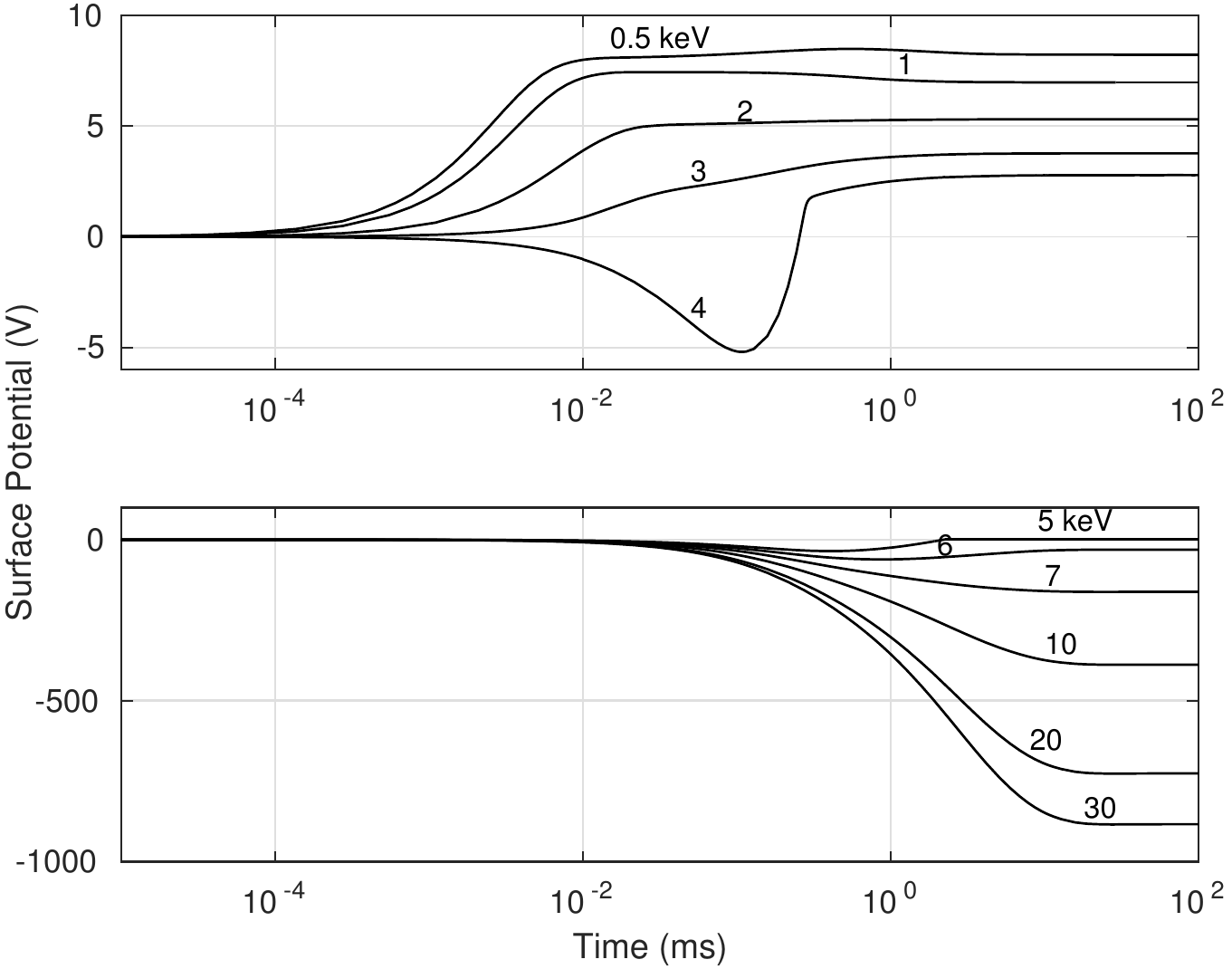}
\caption{Time evolution of the electric potential at the surface of an amorphous
alumina sample versus PE energy for a defocused beam with current density of $10^{-5}$~A/cm$^2$. 
The distance between $\Sigma_1$ and $\Sigma_2$ is $1$~mm.}   
\label{fig:amorAluminaPotential}
\end{figure}

Taking into account that our approach is fundamentally three- not one-dimensional, the results 
presented in Figs.~\ref{fig:amorAluminaYield} and \ref{fig:amorAluminaPotential} 
show general agreement with the Figures~10 and 11 by Touzin et.al \cite{touzin2006electron}, 
especially for the surface potentials at low PE's and the corresponding 
SE emission rates. However, at higher PE energies the accumulated negative 
potential is smaller (lower bounds: $-0.9$~kV in 3D-DDR against $-2.5$~kV in 1D-FD) and 
the yield collapses to unity faster (upper bounds: $\sim 1$~ms in 3D-DDR against 
$\sim 10$~ms in 1D-FD). 
%%%%%%%%%%%%%%%%%%%%%%%%%%%%%%%%%%%%%%%%%%%%%%%%%%%%%%%%%%%%%%%%%%%%%%%%%%%%%%%%%%%%%%%
%%%%%%%%%%%%%%%%%%%%%%%%%%%%%%%%%%%%%%%%%%%%%%%%%%%%%%%%%%%%%%%%%%%%%%%%%%%%%%%%%%%%%%%
%%%%%%%%%%%%%%%%%%%%%%%%%%%%%%%%%%%%%%%%%%%%%%%%%%%%%%%%%%%%%%%%%%%%%%%%%%%%%%%%%%%%%%%

It appears that the distance to the closest Dirichlet boundary,  
where the electric potential is maintained at some 
fixed value, e.g., zero, strongly affects the value of the surface potential
at the sample-vacuum interface.
Apparently, the most important parameter controlling the magnitude of the potential 
is not the total charge density, as one would naively assume, but the proximity 
to an ohmic contact. Most likely this is due to the image-charge effect, which 
partially screens the charge accumulated in the sample.

Things are complicated by the fact that providing at least one Dirichlet 
boundary condition is essential for the numerical stability 
(possibly, existence and uniqueness of the solution as well) of the DDR equations. 
In fact, in the case of an isolated sample, the numerical solution of
our nonlinear problem is only possible through the so-called fully coupled approach, 
since the Dirichlet condition is associated with the Poisson equation,
which, therefore, must remain coupled to the rest of equations during the iterations.
As the boundary conditions at the interfaces of the sample are
not of Dirichlet type, the only available remote surface to impose this type of 
condition is $\Sigma_{1}$. 

\begin{figure}[t!]
\includegraphics[scale=0.62] {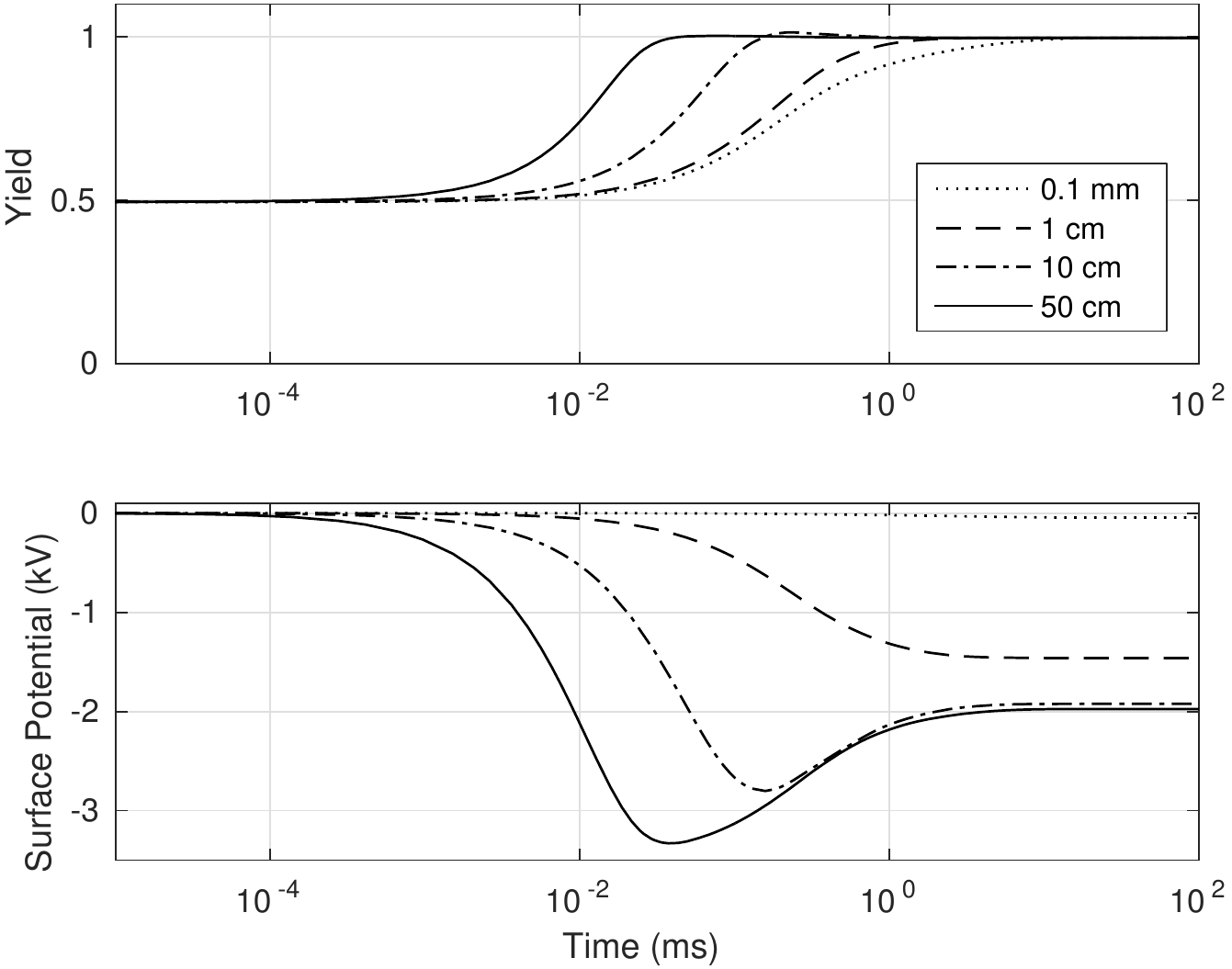}
\caption{Time evolution of the yield and the surface potential from an amorphous
alumina sample at $10$~keV PE energy versus the proximity to the ground contact 
for a for a defocused beam with current density of $10^{-5}$~A/cm$^2$.}
\label{fig:amorAluminaProximity}
\end{figure}

Numerically, the screening effect of the Dirichlet condition can be minimized 
by placing the ohmic contact $\Sigma_{1}$ as far as computationally possible from 
the sample surface $\Sigma_{2}$. 
Thus, we have placed $\Sigma_{1}$ at various distances 
from $\Sigma_{2}$ and, as can be seen in Fig.~\ref{fig:amorAluminaProximity},
the surface potential does reach significant negative values 
when the Dirichlet boundary is far enough. However, the time of collapse of 
yield to unity becomes even shorter in these numerical experiments and remains 
at odds with the previous one-dimensional FD simulations.

%%%%%%%%%%%%%%%%%%%%%%%%%%%%%%%%%%%%%%%%%%%%%%%%%%%%%%%%%%%%%%%%%%%%%%%%%%%%%%%%%%%%%%%
%%%%%%%%%%%%%%%%%%%%%%%%%%%%%%%%%%%%%%%%%%%%%%%%%%%%%%%%%%%%%%%%%%%%%%%%%%%%%%%%%%%%%%%

%%%%%%%%%%%%%%%%%%%%%%%%%%%%%%%%%%%%%%%%%%%%%%%%%%%%%%%%%%%%%
\section{\label{sec:focused}Focused and moving beams}
\begin{figure}[t!]
\includegraphics[scale=0.62] {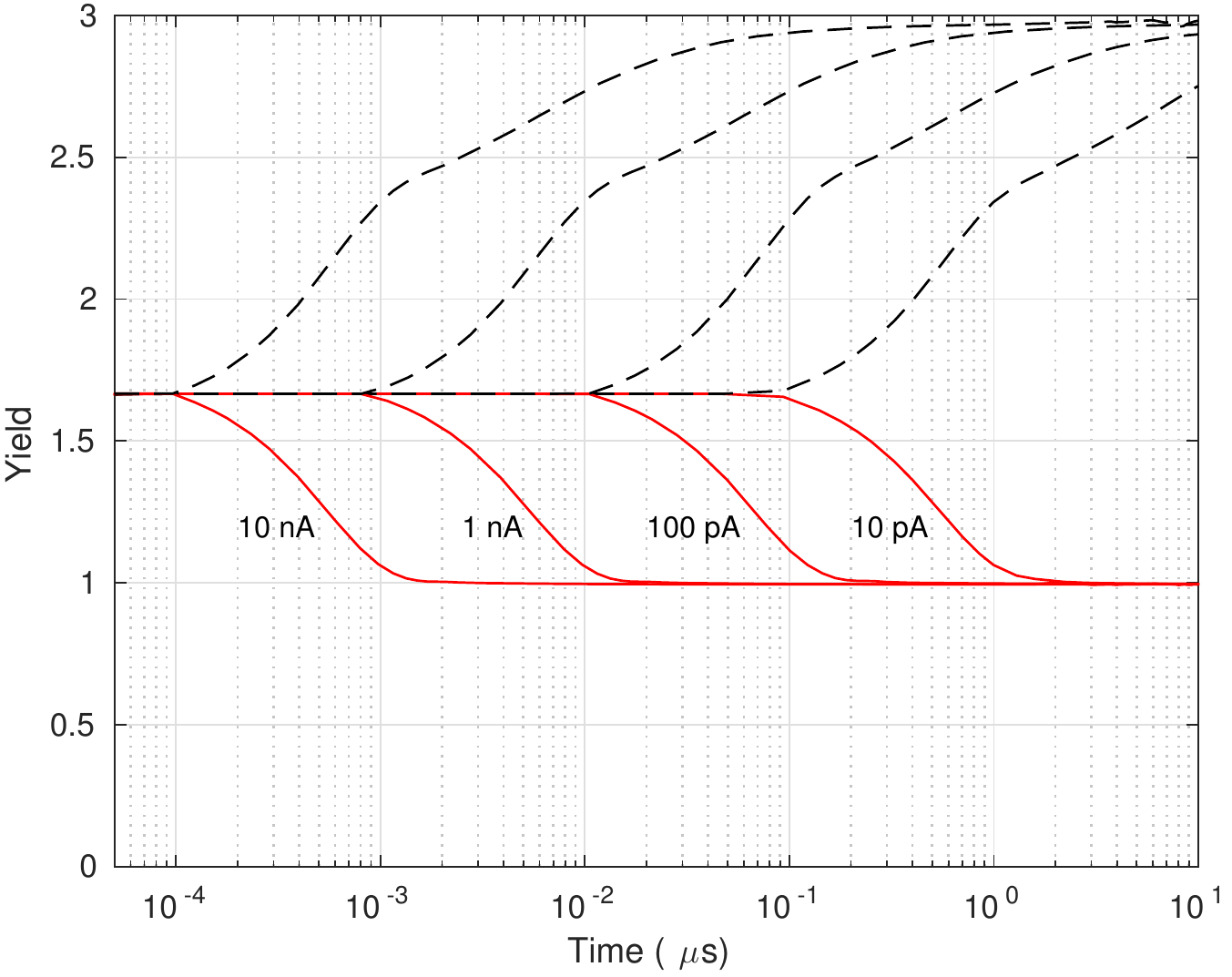}
\includegraphics[scale=0.62] {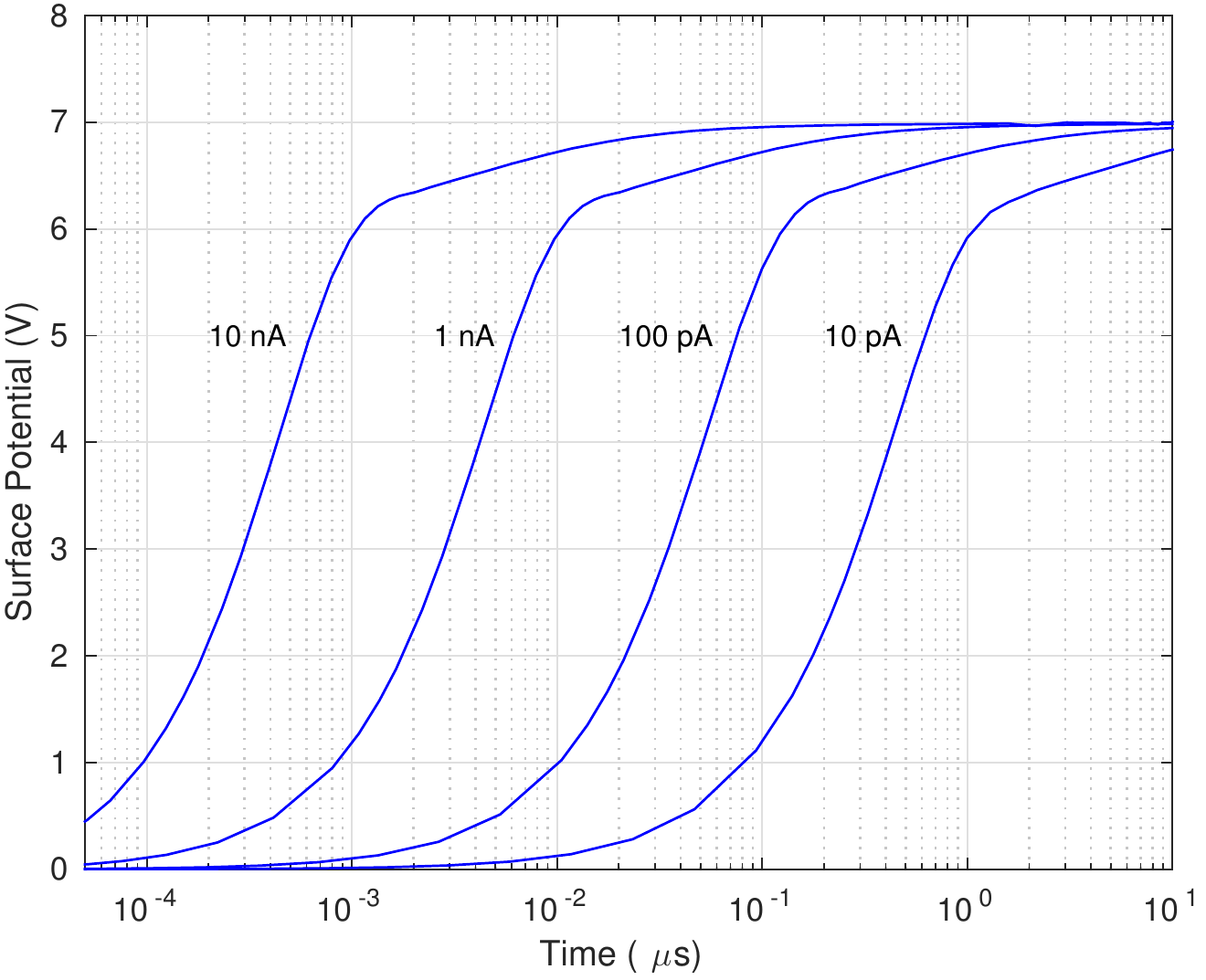}
\caption{Time evolution of the yield (top) and the surface potential at the beam entry 
point (bottom) for a sapphire sample continuously irradiated by a focused stationary 
beam at $5$~keV PE energy -- effect of beam current.
Top: dashed line -- positive part of the SE emission rate through the sample-vacuum interface, 
solid line -- net SE emission rate, including the negative tertiary-electron current.}   
\label{fig:sapphireVersusCurrent}
\end{figure}

\begin{figure}[t!]
\includegraphics[scale=0.62] {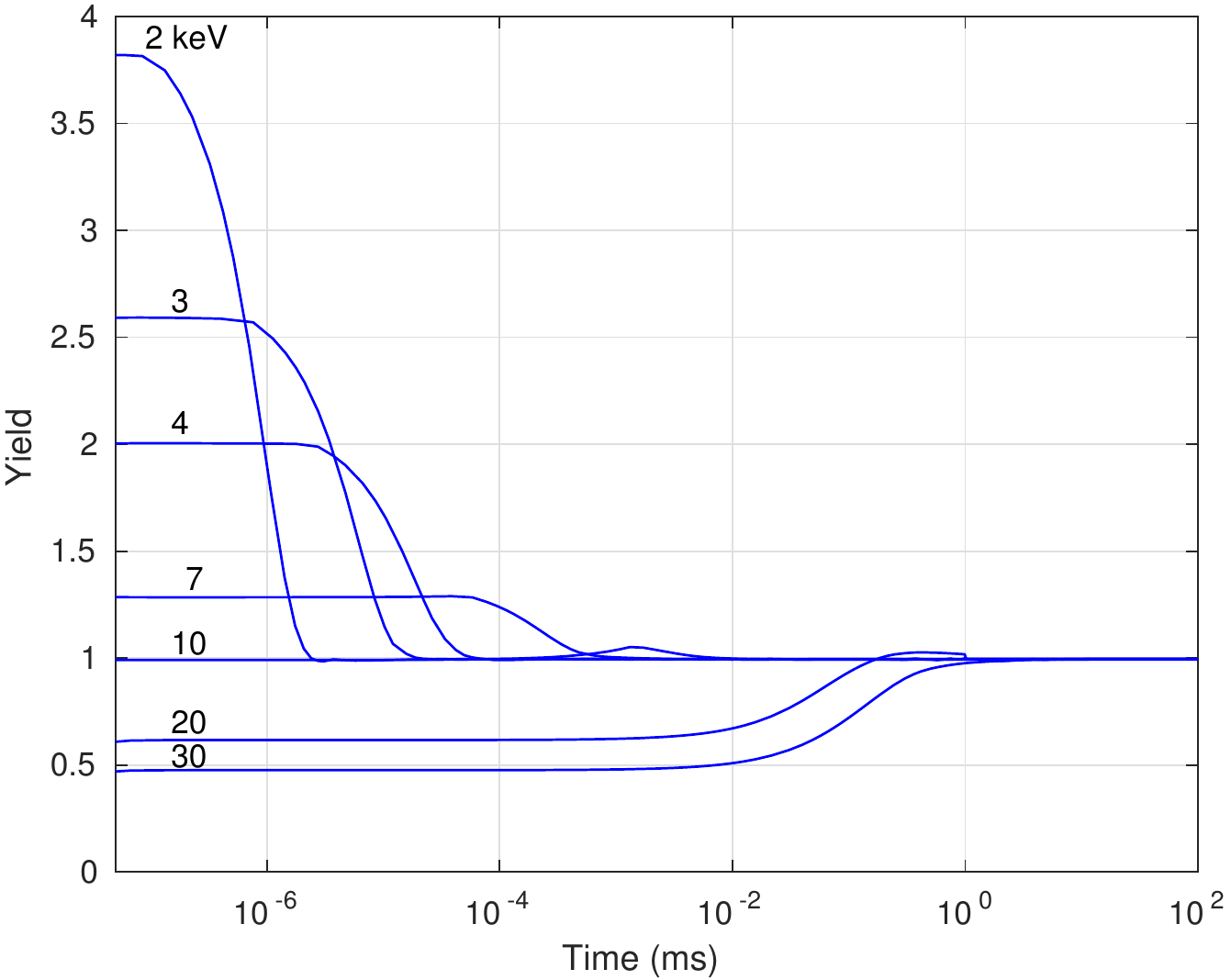}
\includegraphics[scale=0.62] {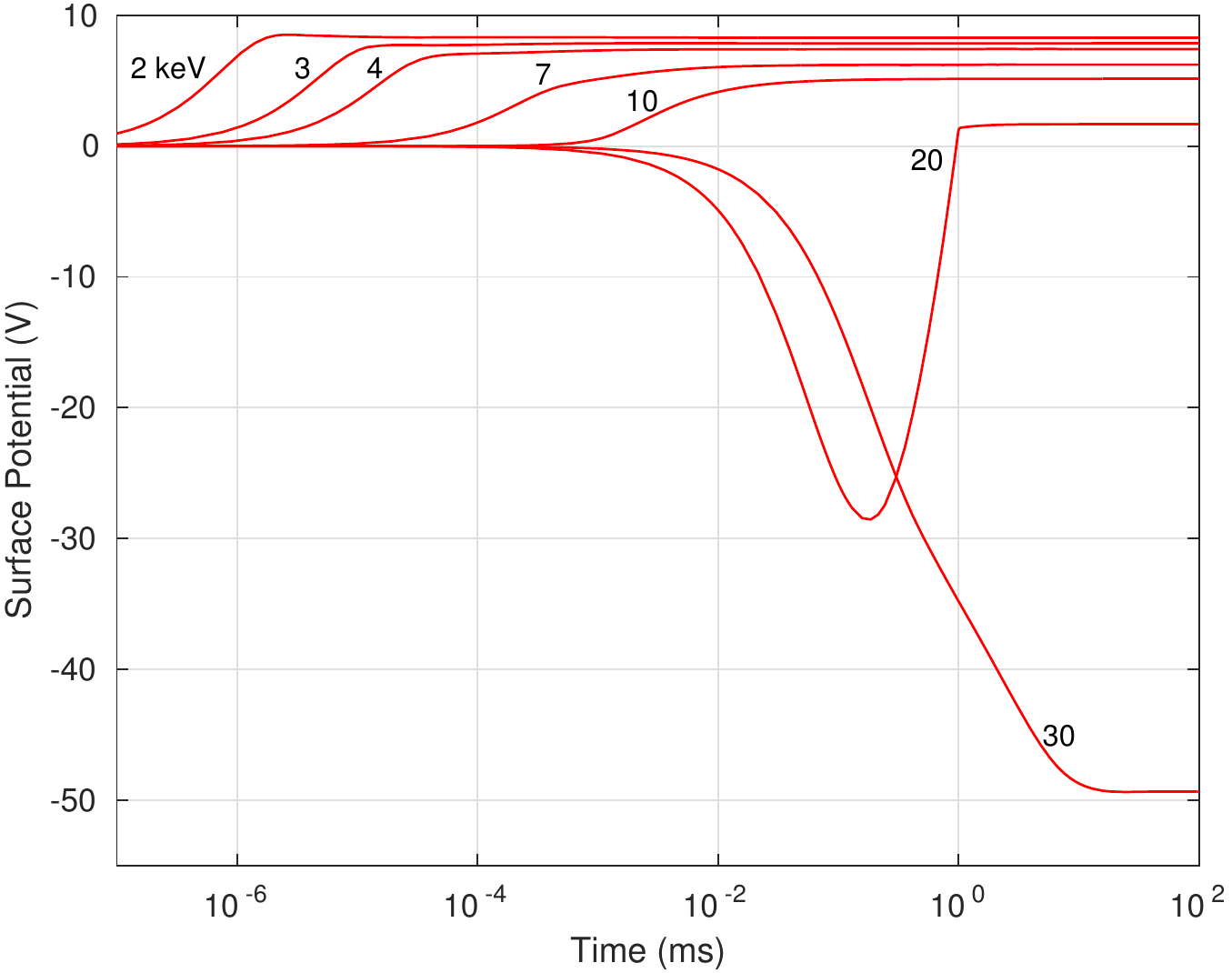}
\caption{Time evolution of the yield (top) and the surface potential at the beam entry 
point (bottom) for a sapphire sample continuously irradiated by a focused stationary 
beam of $100$~pA -- effect of PE energy.}   
\label{fig:sapphireVersusPE}
\end{figure}

\begin{figure}[t!]
\includegraphics[scale=0.62] {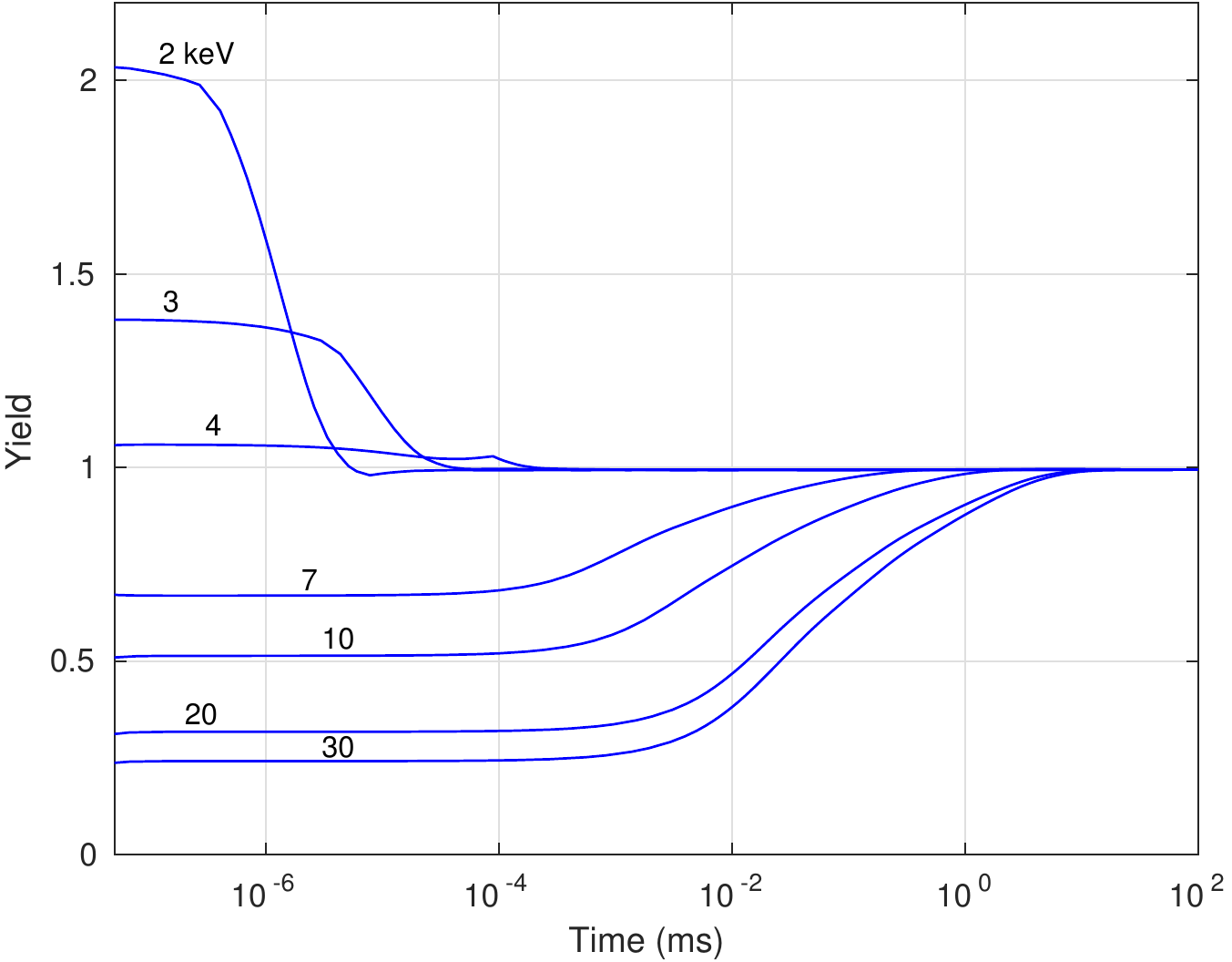}
\includegraphics[scale=0.62] {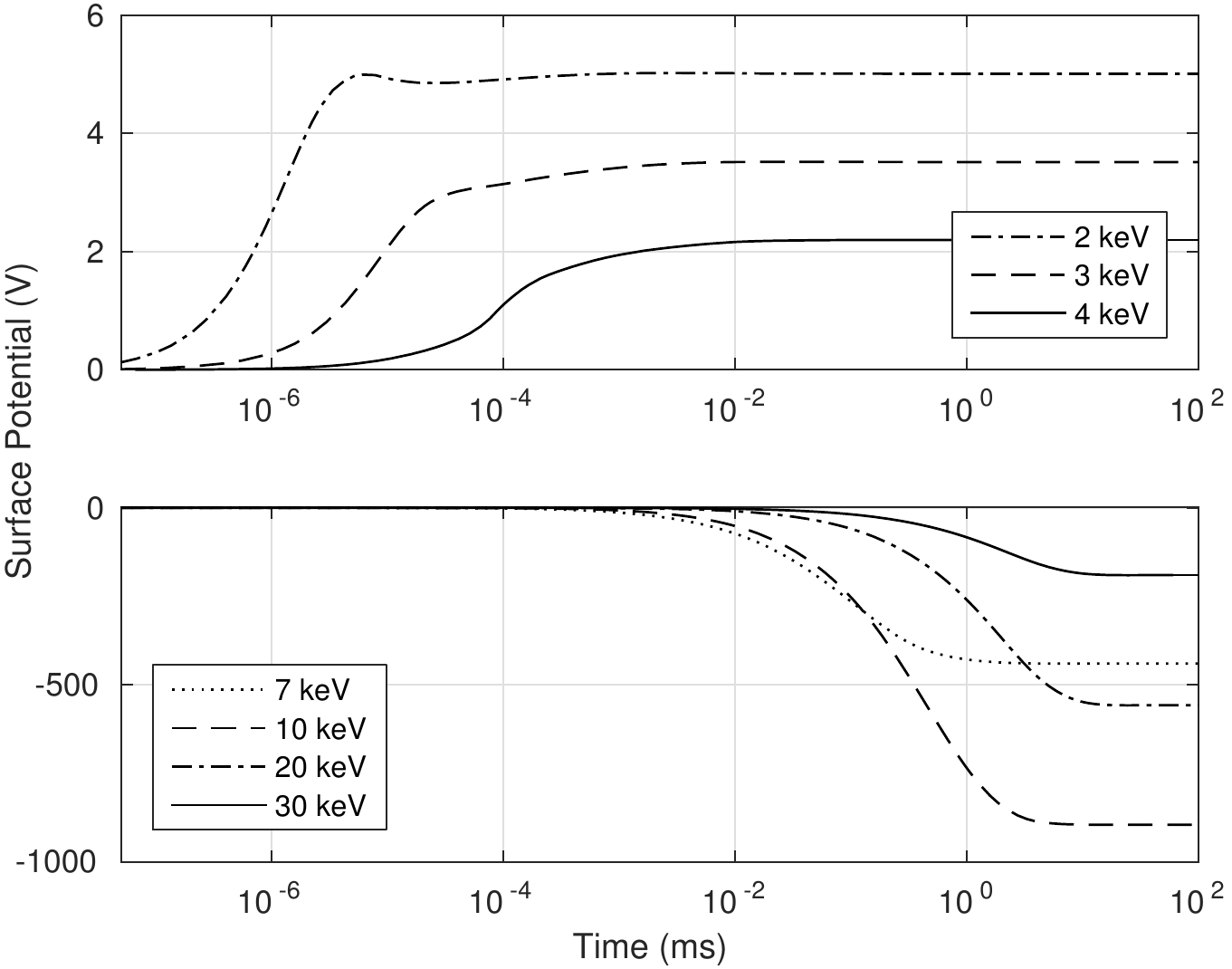}
\caption{Time evolution of the yield (top) and the surface potential at the beam entry 
point (middle and bottom) for a silica sample continuously irradiated by a focused stationary 
beam of $100$~pA -- effect of PE energy.}   
\label{fig:silicaVersusPE}
\end{figure}

%%%%%%%%%%%%%%%%%%%%%%%%%%%%%%%%%%%%%%%%%%%%%%%%%%%%%%%%%%%%%
In this section we simulate sustained irradiation of sapphire, silica, and mixed 
targets by focused stationary and moving electron beams with beam currents typical 
in SEM. In the first series of numerical experiments we
use the axially symmetric target of Fig.~\ref{fig2} illuminated in the 
middle by a focused stationary beam. The distance between $\Sigma_1$ 
and $\Sigma_2$ is set to $0.1$~mm. The samples studied in 
Fig's~\ref{fig:sapphireVersusCurrent}--\ref{fig:silicaVersusPE} are isolated in the 
sense that the only boundary penetrable for particles in the sample-vacuum 
interface $\Sigma_{2}$.
We consider the worst case scenario -- perfect focusing -- where all PE's
hit the same spot on the sample surface. It is easy to deduce that defocusing
will affect low-energy PE's with their small impact zones much stronger 
than higher energy PE's with their extended impact zones. To anticipate 
the results for more realistic partially focused beams the reader is advised to compare
plots of this Section~\ref{sec:focused} with those presented in
Section~\ref{sec:defocused}.

Figure~\ref{fig:sapphireVersusCurrent} pertains to an unpolished sapphire sample irradiated 
at $5$~keV, where the standard yield is around $1.7$ as can be deduced 
from Fig.~\ref{fig:sapphireSingle}~(right). 
The net SE emission rate -- yield for short -- starts at the standard yield value,
but after a certain interval of time drops to unity for all beam currents. 
The stronger the current, the shorter is the standard yield interval preceding the drop.
In fact, it is easy to calculate that the drop in the yield happens after a certain 
amount of charge has been injected into the sample by the beam, which confirms
conclusions of many previous investigations. The point of fastest decline in the 
yield roughly corresponds to $3~\times~10^{-18}$~C of injected primary charge, i.e.,
approximately $19$ primary electrons.

%%%%%%%%%%%%%%%%%%%%%%%%%%%%%%%%%%%%%%%%%%%%%%%%%%%%%%%%%%%%%%%%%%%%%%%%%%%%%%%%%%%%%%%
%%%%%%%%%%%%%%%%%%%%%%%%%%%%%%%%%%%%%%%%%%%%%%%%%%%%%%%%%%%%%%%%%%%%%%%%%%%%%%%%%%%%%%%

The DDR model does not substantiate the usual intuitive explanation\cite{renoud2004secondary,touzin2006electron} 
concerning the reasons behind this seemingly inevitable convergence of yield 
to unity with time. 
Commonly it is argued that the charging of the sample leads to the change in the
landing energy of PE's, so that the yield no longer corresponds to the standard
yield of that energy, but rather to another point on the standard yield curve of 
Fig.~\ref{fig:sapphireSingle}~(right).
If, for example, the standard yield is greater than one, then the sample accumulates
positive charge. The landing energy increases and one should look to the 
right along the standard-yield curve to know what the new yield should be. 
If, on the other hand, the standard yield is less than one, then the accumulated 
negative charge reduces the landing energy of the PE's, thus, moving to the left along 
the standard-yield curve. Thus, it is argued, a yield larger than one
would eventually lead to a positive potential high enough to shift the landing energy
of primary electrons to the second unity-crossing point on the standard yield curve.
This argument, while intuitively appealing, does not take into account the 
spatial distribution, the dynamics, and the screening of charges. 
In fact, in our simulations the accumulated potential was never strong enough for the 
landing energy to reach a unity-crossing point.

For example, Figure~\ref{fig:sapphireVersusCurrent}~(bottom) clearly shows that the value of the
positive surface potential is insignificant with respect to the PE energy and cannot
possibly change the landing energy by so much that it becomes $10$~keV -- the second 
point along the standard-yield curve where it crosses the unity line. What the DDR model
shows, though, is that the drop in the yield coincides with the rapid increase in the
tertiary current, caused by the relatively weak positive surface potential attracting 
low-energy SE's back to the sample. 
Figure~\ref{fig:sapphireVersusCurrent}~(top) compares the contribution of the positive part 
$v_{e}(n-n_{i})$ of the emission current density (dashed lines) to the net SE emission 
rate (solid lines). The onset of the tertiary current can be deduced from the emergent discrepancy
between the solid and dashed curves, which coincides with the positive surface potential 
reaching the value $V_{min}=1$~V in the bottom plot of Fig.~\ref{fig:sapphireVersusCurrent}. 
Moreover, tertiary current remains significant even after the net 
yield reaches unity. Thus, the unity yield is the product of a neat dynamic balance between 
the PE injection, positive outward SE emission, and the reverse tertiary current. 
The result is a steady-state process and the conservation of total charge (on average): 
one PE in, one SE out, and a conserved `circular' current at the sample-vacuum interface.

Figure \ref{fig:sapphireVersusPE} (sapphire) and Figure~\ref{fig:silicaVersusPE} (silica) 
correspond to the beam current of $100$~pA and show the time evolution of the yield 
and potential for various PE energies. Comparison with the defocused beam irradiation
of Fig's.~\ref{fig:amorAluminaYield}--\ref{fig:amorAluminaPotential} reveals a
larger discrepancy in convergence times of the yield to unity for different 
energies in the focused beam case. The yield drops much sooner at lower PE energies
than it increases at higher PE energies.
%%%%%%%%%%%%%%%%%%%%%%%%%%%%%%%%%%%%%%%%%%%%%%%%%%%%%%%%%%%%%%%%%%%%%%%%%%%%%%%%%%%%%%%
%%%%%%%%%%%%%%%%%%%%%%%%%%%%%%%%%%%%%%%%%%%%%%%%%%%%%%%%%%%%%%%%%%%%%%%%%%%%%%%%%%%%%%%

%%%%%%%%%%%%%%%%%%%%%%%%%%%%%%%%%%%%%%%%%%%%%%%%%%%%%%%%%%%%%%%%%%%%%%%%%%%%%%%%%%%%%%%
%%%%%%%%%%%%%%%%%%%%%%%%%%%%%%%%%%%%%%%%%%%%%%%%%%%%%%%%%%%%%%%%%%%%%%%%%%%%%%%%%%%%%%%

Similarly, from the surface potential plots of Fig's~ \ref{fig:sapphireVersusPE}--\ref{fig:silicaVersusPE} 
we conclude that the rise of sub-unit 
yields (above $10$~keV for sapphire and above $4$~keV for silica) to unity  
cannot be explained by the change in the landing energy, as the associated 
potential is never negative enough for that.
Minimizing the screening by removing the Dirichlet boundary $\Sigma_{1}$ 
farther away from the sample-vacuum interface $\Sigma_{2}$
we could bring the surface potential in silica down to 
$-15$~kV, which, however, was still not enough 
to decrease the landing energy of PE's from $30$~keV down to the required 
$4.35$~keV, where the standard yield of silica is equal to one.
We propose a much simpler alternative explanation: sub-unit yields increase the number of
free electrons near the sample-vacuum interface, which, in its turn, 
increases the SE emission rate up until the steady-state condition of unit yield is reached.
Sometimes, as at $20$~keV in sapphire and at $4$~keV in amorphous alumina, 
the yield grows so fast that there is an overshoot, and it
temporarily becomes larger than one, causing a positive surface potential,
which creates the tertiary current pulling the yield back to unity. 

\begin{figure}[t!]
\includegraphics[scale=0.62] {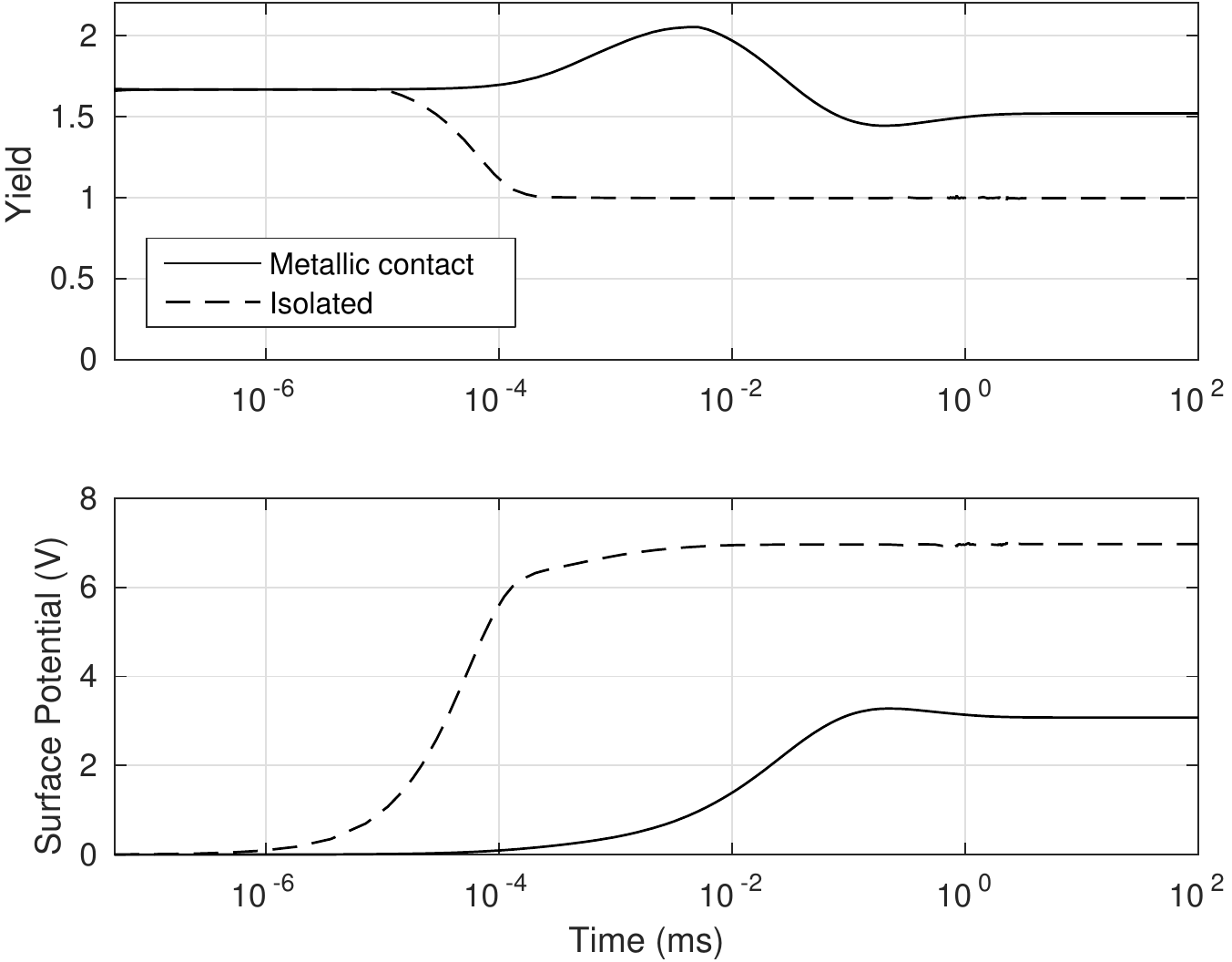}
\caption{Time evolution of the yield and the surface potential from 
isolated and grounded sapphire 
samples for $5$~keV PE energy and $100$~pA beam current. 
The distance between $\Sigma_1$ and $\Sigma_2$ is $0.1$~mm.}   
\label{fig:sapphireMetalVSisolated}
\end{figure}

The yield does not always have to drop/increase to unity, though.
If it was the case, all insulators would look exactly the same under SEM. 
One possible scenario, where the 
yield may not converge to unity, is a (partially) grounded sample.
The condition on charge conservation that requires a unit yield in an isolated sample
may be relaxed if the sample is grounded. It is, of course, an open question
whether a contact between an insulator and, say, a metallic grounded 
holder can ever be made efficient enough to allow for an easy passage 
of charges. Assuming for simplicity a perfect ohmic contact,
the charge conservation no longer requires the exact unit yield
for the sample-vacuum interface as additional electrons may 
enter the sample via the ground channel. This situation is illustrated
in Fig.~\ref{fig:sapphireMetalVSisolated}, where we have imposed
an ohmic boundary condition on the side of sapphire sample.
Although the yield in such a grounded sample does not stay at the level of the 
standard yield at that energy, after a few oscillations it
stabilizes at a slightly lower value, well above unity. 
This effect is also observed in samples with a relatively
poor ground contact described by a Robin-type boundary condition with
a low surface recombination velocity.

%%%%%%%%%%%%%%%%%%%%%%%%%%%%%%%%%%%%%%%%%%%%%%%%%%%%%%%%%%%%%%%%%%%%%%%%%%%%%%%%%%%%%%%
%%%%%%%%%%%%%%%%%%%%%%%%%%%%%%%%%%%%%%%%%%%%%%%%%%%%%%%%%%%%%%%%%%%%%%%%%%%%%%%%%%%%%%%

\begin{figure*}[t!]
\includegraphics[scale=0.9] {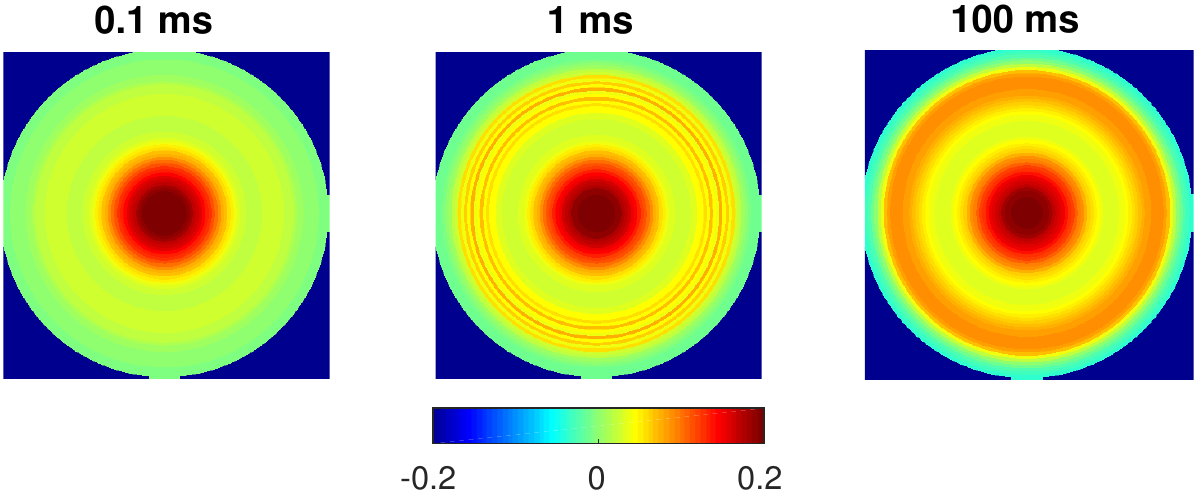}
\includegraphics[scale=0.9] {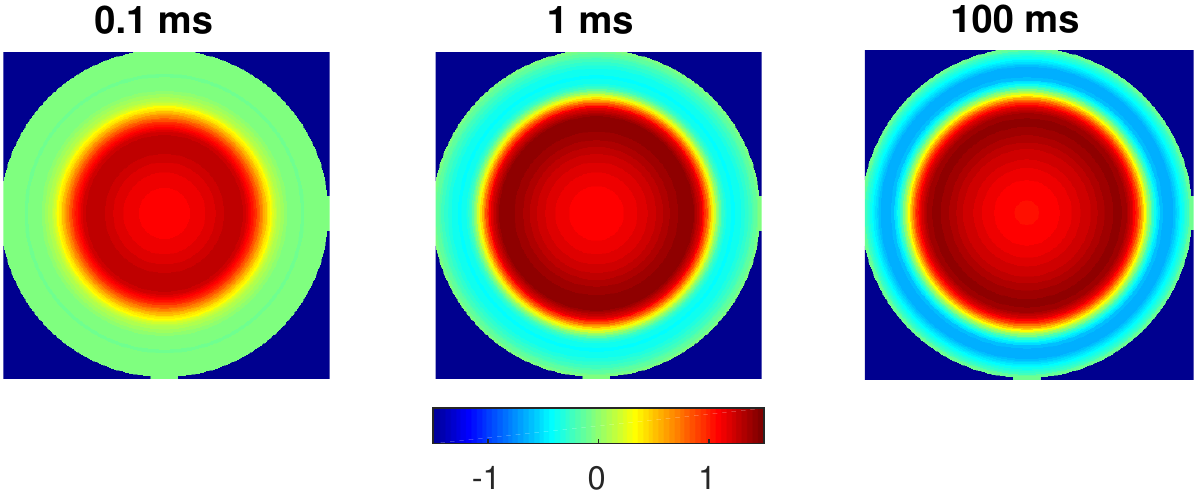}
\caption{Build up of the surface charge $(C+p-n-n_T)q~(\text{C/cm}^3)$ 
in isolated (top row) and grounded (bottom row) sapphire 
samples irradiated by a focused stationary $100$~pA beam of $5$~keV PE's. 
The distance between $\Sigma_1$ and $\Sigma_2$ is $0.1$~mm.}   
\label{fig:surfaceChargeSapphire}
\end{figure*}

Although, the surface potential does take longer to build up in a sample 
with contact, Fig.~\ref{fig:sapphireMetalVSisolated}~(bottom), the behavior of 
the surface potential at the injection point is not very
revealing. It is, perhaps, more instructive to look at the
distribution of the total charge at the surfaces of isolated and grounded samples
under identical irradiation conditions.
While the surface potential is weaker in the grounded case, 
Fig.~\ref{fig:sapphireMetalVSisolated}~(bottom), 
the images of Fig.~\ref{fig:surfaceChargeSapphire} explicitly show that 
the amount of accumulated positive charge at the surface of a grounded sample 
is higher. 
Also the spatial distributions of the surface charge are different. 
A large disk of positive charge surrounded by a ring of negative charge 
is seen in the grounded sample, whereas, in the isolated sample 
most of the positive surface charge is concentrated around the injection point followed
by a weaker positive ring some distance away. 

%%%%%%%%%%%%%%%%%%%%%%%%%%%%%%%%%%%%%%%%%%%%%%%%%%%%%%%%%%%%%%%%%%%%%%%%%%%%%%%%%%%%%%%
%%%%%%%%%%%%%%%%%%%%%%%%%%%%%%%%%%%%%%%%%%%%%%%%%%%%%%%%%%%%%%%%%%%%%%%%%%%%%%%%%%%%%%%

Another situation well-known to SEM practitioners where the yield does not drop/increase 
to unity is the rapid scanning of the sample by a moving focused beam.
To simulate the scanning process the source function (\ref{equ:Gus}) 
has to be modified to account for the motion of the beam. 
This is achieved by setting ${\mathbf x}_{0}(t)={\mathbf x}_{0}+{\mathbf v} t$,
where ${\mathbf v}$ is the velocity of beam displacement in the horizontal plane.
Consider a $1\times 1$~$\mu\text{m}^{2}$ sample surface imaged with a $1000 \times 1000$ 
pixels resolution at the rate of $30$ frames per second. Then, the beam moves 
across the sample with the horizontal speed 
$\vert{\mathbf v}\vert\approx $~$33$~$\mu$m/s. 

\begin{figure}[t!]
\begin{tikzpicture}
     \draw[gray!100!white] (0,0) rectangle (4,4);
     \draw (4,0)  -- (5,1);
     \draw (4,4)  -- (5,5);
     \draw (5,1)  -- (5,5);
     \draw (5,5)  -- (1,5);
     \draw (1,5)  -- (0,4);
     \draw [dashed] (1,5)  -- (1,1);
     \draw [dashed] (1,1)  -- (0,0);
     \draw [dashed] (1,1)  -- (5,1);
     \draw  (2,0)  -- (2,4);
     \draw  (2,4)  -- (3,5);
     \draw [dashed] (3,5)  -- (3,1);
     \draw [dashed] (3,1)  -- (2,0);
     \node[] at (1.5,0.5) {$\text{Al}_2\text{O}_3$};
     \node[] at (3.5,0.5) {$\text{Si}\text{O}_2$};
     \draw [red,thick,dashdotted] (0.5,4.5)  -- (4.5,4.5);
     \draw [thick,->] (1.5,5.5) -- (1.5,4.5);
     \node[] at (1.5,5.7) {Scan line};
\end{tikzpicture}
\caption{An inhomogeneous sample consisting of vertically stacked sapphire 
and silica blocks.}
\label{fig:inhomogeneous}
\end{figure}
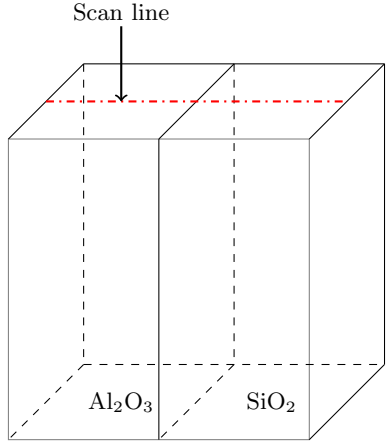

We consider an inhomogeneous sample consisting of adjacent blocks of 
sapphire and silica, see Fig.~\ref{fig:inhomogeneous}.
Samples consisting of one insulator on top of another have been previously studied 
with a one-dimensional approach \cite{cornet2008electron}, 
while vertical stacks of insulators, similar to the one considered here, 
have been recently investigated experimentally \cite{kim2010charging}. 
%%%%%%%%%%%%%%%%%%%%%%%%%%%%%%%%%%%%%%%%%%%%%%%%%%%%%%%%%%%%%%%%%%%%%%%%%%%%%%%%%%%%%%%
%%%%%%%%%%%%%%%%%%%%%%%%%%%%%%%%%%%%%%%%%%%%%%%%%%%%%%%%%%%%%%%%%%%%%%%%%%%%%%%%%%%%%%%

We simulate a single scan line through the middle of the sample 
perpendicular to the interface between the adjacent insulators. Across the vertical 
interface between the two different insulators the source 
function (\ref{equ:Gus}) exhibits a discontinuity due to the change in material
density and the corresponding maximum PE penetration depth.

\begin{figure}[t!]
\includegraphics[scale=0.64] {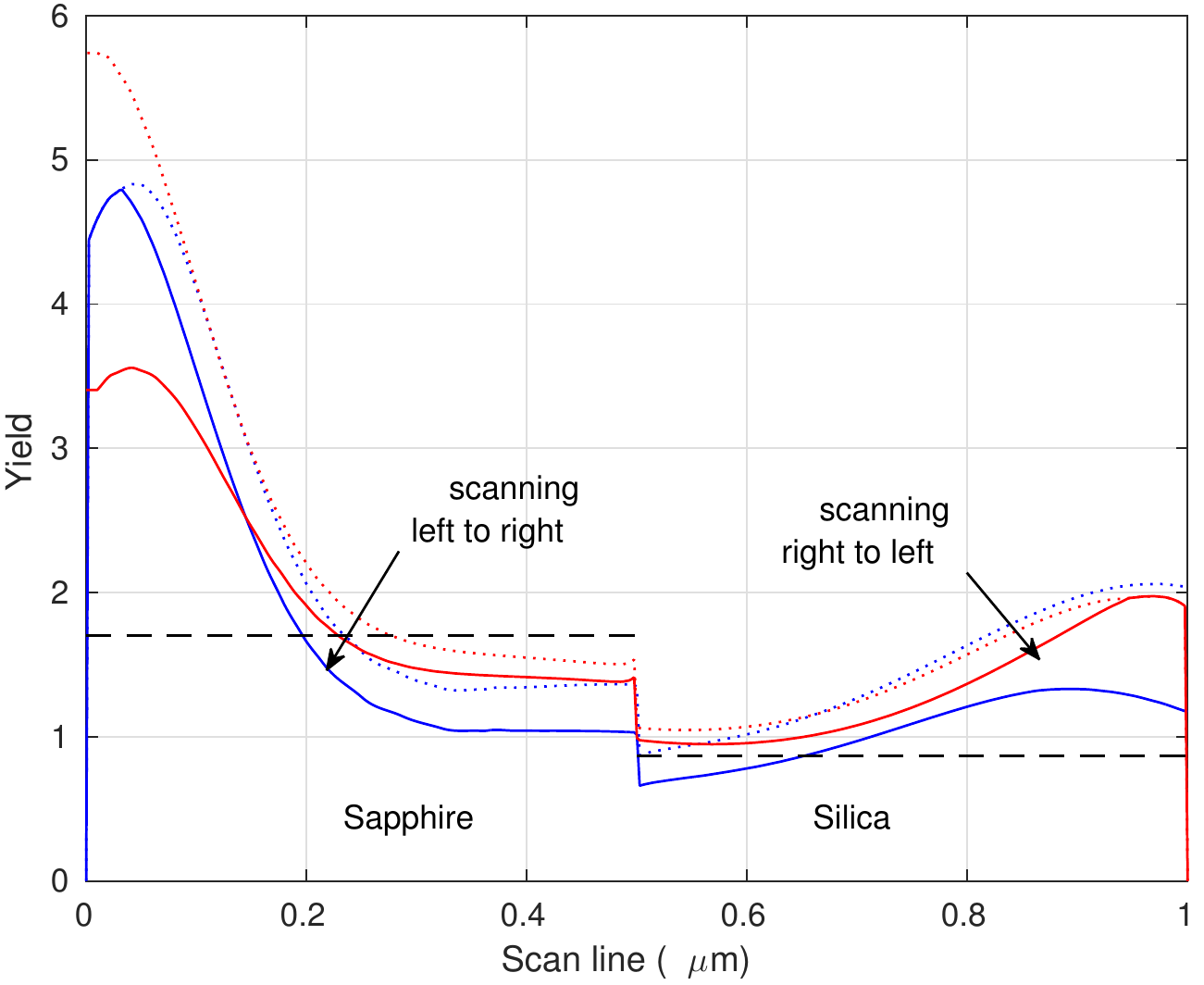}
\caption{Yield as a function of beam position while 
scanning an inhomogeneous sapphire-silica sample ($5$~keV, $10$~pA). 
Solid lines -- SE emission rate; dotted -- positive part of the emission current;
dashed -- standard yield.}   
\label{fig:sapphireSilicaYield}
\end{figure}

Since cylindrical symmetry is lost, the following
DDR computations had been performed in the full three-dimensional mode. 
Figure~\ref{fig:sapphireSilicaYield} shows the yield as a function of the beam position
along its trajectory for an isolated sample. These curves correspond to the 
intensity of pixels in a single-line SEM image. 
The standard yields of both insulators at the considered PE energy 
are also shown as dotted lines.

%%%%%%%%%%%%%%%%%%%%%%%%%%%%%%%%%%%%%%%%%%%%%%%%%%%%%%%%%%%%%%%%%%%%%%%%%%%%%%%%%%%%%%%
\begin{figure*}[t!]
\includegraphics[scale=1] {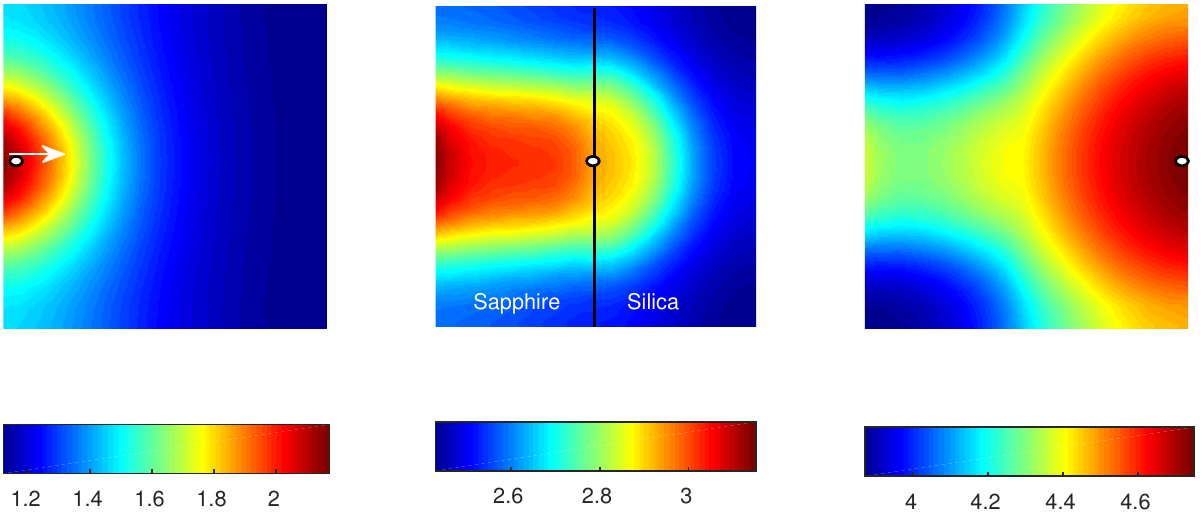}
\includegraphics[scale=1] {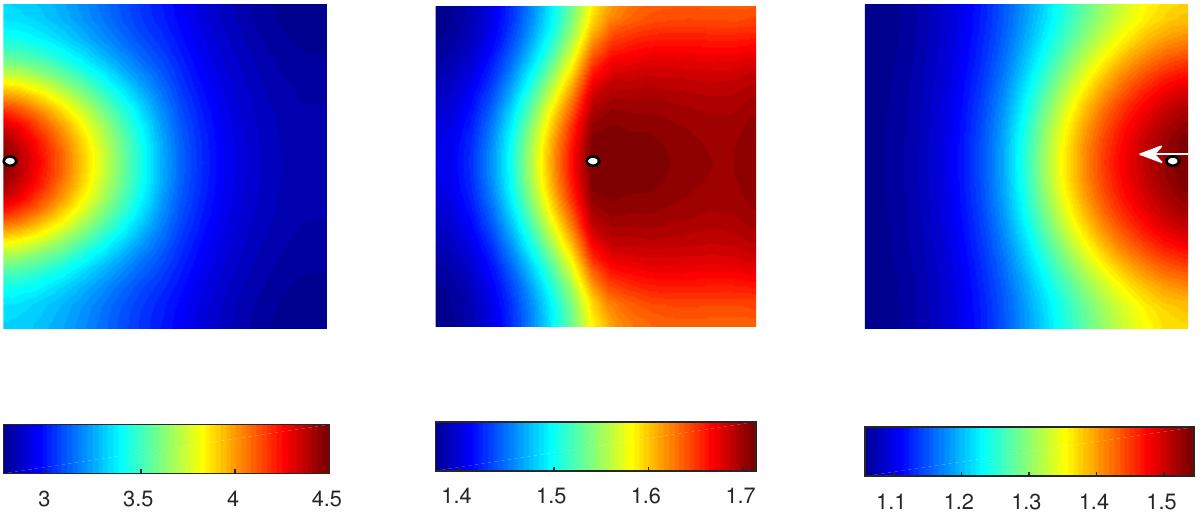}
\caption{Surface potential at three beam locations (white circles) during sample
scanning ($5$~keV, $10$~pA). 
Top and bottom rows correspond to opposite scanning directions (indicated by arrows).}   
\label{fig:surfacePotential}
\end{figure*}
%%%%%%%%%%%%%%%%%%%%%%%%%%%%%%%%%%%%%%%%%%%%%%%%%%%%%%%%%%%%%%%%%%%%%%%%%%%%%%%%%%%%%%%

First of all we notice the difference between the left-to-right 
(from sapphire to silica) and the right-to-left (from silica to sapphire)
scanning modes. This difference is easy to understand by looking at 
Fig.~\ref{fig:surfacePotential} where the images show the surface potential at 
the same beam locations during these two scans. Since the charging
of sapphire is stronger than that of silica, the resulting residual charge
strongly depends on the scan history.

Otherwise, the scans of Fig.~\ref{fig:sapphireSilicaYield} have several common 
features. One can notice higher yields in the neighborhood of the
sample edges due to increased emission via the vertical interfaces. 
This is a well-known effect -- the sample edges look brighter in SEM images
compared to the rest of the sample surface.
One can also see the drop of the yield to unity during left-to-right scanning 
due to continuous charging of the sapphire part. This charging also causes the yield in silica 
part to drop below its standard value. The relatively smaller charging during 
the right-to-left scanning does not allow the yield in silica to reach its 
standard below-unity value after the initial edge-related surge, and keeps the 
yield below the standard value when the beam crosses into the sapphire part.
Additional simulations show that reducing the
beam current (down to a few pA) while maintaining a high beam displacement velocity gives scans that  
truthfully reflect the standard yields of each part of the sample. Unfortunately,
in practice this would, probably, result in a bad signal to noise ratio.

%%%%%%%%%%%%%%%%%%%%%%%%%%%%%%%%%%%%%%%%%%%%%%%%%%%%%%%%%%%%%
\section{Conclusions}
%%%%%%%%%%%%%%%%%%%%%%%%%%%%%%%%%%%%%%%%%%%%%%%%%%%%%%%%%%%%%
The self-consistent DDR method proposed in \cite{raftari2015self} 
has been substantially modified in the present paper to include the 
dynamic trap-assisted generation-recombination model and a novel self-consistent 
boundary condition accounting for tertiary electrons. 
The method has been calibrated against experimental 
data do deliver exact standard yields for alumina and silica samples 
over a large range of PE energies. For alumina and silica all calibrated 
parameters remain within or close to their reported uncertainty bounds, thereby
further confirming the acceptability of model approximations. 
Time-domain 
simulations with defocused beams have been compared to the previously 
published results from a one-dimensional Flight-Drift model 
demonstrating similar long-time behavior. 
Our investigations so far show that the 
initial high-energy transport stage can, indeed, be approximated
by a semi-empirical source function and low-energy material parameters, 
whereas, subsequent transport stages fall within the original domain of 
validity of the low-energy DDR method.

Simulations with stationary focused beams confirm that in electrically isolated 
samples the yield collapses to unity after a certain number of primary electrons,
which depends on the PE energy, has been injected roughly at the same location 
on the sample surface.
However, our simulations do not support the widespread intuitive explanation of 
this phenomenon in terms of the changing landing energy of PE's. The effect appears
to have dynamic origins and is related to tertiary currents and 
transient changes in the distribution of charges close to the sample surface.

The surface potential is strongly affected by the proximity of metallic grounded 
surfaces due to the associated charge screening. This may lead to 
misinterpretation of charging effects, if one relies solely on the surface potential 
measurements, but also may present an opportunity to alleviate SEM image distortions. 
Our simulations show that a good ground contact could also prevent the collapse 
of the yield to unity.

We have presented, probably, the first 3D simulations of a laterally inhomogeneous sample
irradiated by a moving beam that take into account 
both the dynamic charge trapping/de-trapping and the tertiary electrons. 
While to a certain extent the yield obtained during 
this realistic simulations could be interpreted on the basis of 
time-domain results with stationary beams, some effects are unique to dynamic scanning. 
For example, the scan profile appears to depend on the direction of 
scanning.

A recent review by Walker et al \cite{walker2016simulations} mentions the 
lack of reliable simulations related to low-energy SEM studies. 
We hope to partly fill this gap with the present modified and calibrated 
version of the DDR method. Our method as well as other simulation software would
greatly benefit from publicly available high-quality time-domain data 
in addition to the already available standard yields. While we realize 
that direct time-domain sampling of detector currents may be difficult, 
it should be possible to collect and publish single line scans of $\sim 1~\mu\text{m}$
insulator targets for a rage of dwell times (scan speeds).

%%%%%%%%%%%%%%%%%%%%%%%%%%%%%%%%%%%%%%%%%%%%%%%%%%%%%%%%%%%%%
\section*{Acknowledgements}
%%%%%%%%%%%%%%%%%%%%%%%%%%%%%%%%%%%%%%%%%%%%%%%%%%%%%%%%%%%%%
The authors acknowledge the partial financial support of the Thermo Fisher Scientific 
(formerly FEI, Netherlands)
and many fruitful discussions with H.~van~der~Graaf, C.W.~Hagen, S.~Tao, 
A.M.M.G.~Theulings, and H.-W.~Chan 
(all from Charged Particle Optics group at Delft University of Technology). 

%%%%%%%%%%%%%%%%%%%%%%%%%%%%%%%%%%%%%%%%%%%%%%%%%%%%%%%%%%%%%

%***********************************************************************************


\begin{thebibliography}{10}
\bibitem{bass1998measurements}
AD~Bass, P~Cloutier and L~Sanche.
\newblock Measurements of charge accumulation induced by monochromatic low-energy electrons at the surface of insulating samples.
\newblock {\em Journal of applied physics},
84(5), 2740--2748, 1998.
%
\bibitem{vdGraaf2017}
H. van der Graaf, H. Akhtar, N. Budko, H.W. Chan, C.W. Hagen, C.C.T. Hansson, G. N{\"u}tzel, S.D. Pinto, V. Prodanovi{\'c}, B. Raftari, 
P.M. Sarro, J. Sinsheimer, J. Smedley, S. Tao, A.M.M.G. Theulings, C. Vuik.
\newblock The Tynode: a new vacuum electron multiplier. 
\newblock {\em Nuclear Instruments and Methods in Physics Research A} 
847, 148--161, 2017.
%
\bibitem{lee2007layer}
JS~Lee and J~Cho and C~Lee and I~Kim and J~Park and YM~Kim and H~Shin and J~Lee and F~Caruso.
\newblock Layer-by-layer assembled charge-trap memory devices with adjustable electronic properties.
\newblock {\em Nature nanotechnology},
2(12), 790--795, 2007.
%
\bibitem{kulwicki1970diffusion}
BM~Kulwicki and AJ~Purdes.
\newblock Diffusion potentials in BaTiO3 and the theory of PTC materials.
\newblock {\em Ferroelectrics},
1(1), 253--263, 1970.
%
\bibitem{vampola1985aerospacer}
Vampola, AL and Mizera, PF and Koons, HC and Fennell, JF and Hall, DF.
\newblock {\em The Aerospace Spacecraft Charging Document}.
\newblock No. SD-TR-0084A (5940-05)-10. AEROSPACE CORP EL SEGUNDO CA LAB OPERATIONS, 1985.
%
\bibitem{gross1957irradiation}
B~Gross.
\newblock Irradiation effects in borosilicate glass.
\newblock {\em Physical Review}, 107(2), 368--374, 1957.
%
\bibitem{gross1958irradiation}
B~Gross.
\newblock Irradiation effects in Plexiglas.
\newblock {\em Journal of Polymer Science}, 27(115), 135--143, 1958.
%
\bibitem{gross1967high}
B~Gross and S.V~Nablo.
\newblock High Potentials in Electron-Irradiated Dielectrics.
\newblock {\em Journal of Applied Physics}, 38(5), 2272--2275, 1967.
%
\bibitem{gross1973charge}
B~Gross, J~Dow and S.V~Nablo.
\newblock Charge buildup in electron-irradiated dielectrics.
\newblock {\em Journal of Applied Physics}, 44(6), 2459--2463, 1973.
%
%
\bibitem{gross1974charge}
B~Gross,GM~Sessler and JE~West.
\newblock Charge dynamics for electron-irradiated polymer-foil electrets.
\newblock {\em Journal of Applied Physics}, 45(7), 2841--2851, 1974.
%
\bibitem{gross1974transport}
B~Gross and LN de Oliveira.
\newblock Transport of excess charge in electron-irradiated dielectrics.
\newblock {\em Journal of Applied Physics}, 45(11), 4724--4729, 1974.
%
%
\bibitem{le1989study}
R~Le Bihan.
\newblock Study of ferroelectric and ferroelastic domain structures by scanning electron microscopy.
\newblock {\em Ferroelectrics},
97(1), 19--46, 1989.
%
\bibitem{liebl1980sims}
H~Liebl.
\newblock SIMS instrumentation and imaging techniques.
\newblock {\em Scanning},
3(2), 79--89, 1980.
%
\bibitem{meyza2003secondary}
X~Meyza, D~Goeuriot, C~Guerret-Pi{\'e}court, D~Tr{\'e}heux and H-J~Fitting.
\newblock Secondary electron emission and self-consistent charge transport and storage in bulk insulators: Application to alumina.
\newblock {\em Journal of applied physics},
94(8), 5384--5392, 2003.
%
\bibitem{touzin2006electron}
M~Touzin, D~Goeuriot, C~Guerret-Pi{\'e}court, D~Juv{\'e}, D~Tr{\'e}heux and H-J~Fitting.
\newblock Electron beam charging of insulators: A self-consistent flight-drift model.
\newblock {\em Journal of applied physics},
99(11), 114110, 2006.
%
\bibitem{walker2016simulations}
GHC~Walker, L~Frank and I~M{\"u}llerov{\'a}.
\newblock Simulations and measurements in scanning electron microscopes at low electron energy.
\newblock {\em Scanning},
9999, 1--17, 2016.
%
\bibitem{li2011self}
Wei-Qin Li, Kun Mu and Rong-Hou Xia.
\newblock Self-consistent charging in dielectric films under defocused electron beam irradiation.
\newblock {\em Micron},
42(5), 443--448, 2011.
%
\bibitem{li2010surface}
Wei-Qin~Li and Hai-Bo~Zhang.
\newblock The surface potential of insulating thin films negatively charged by a low-energy focused electron beam.
\newblock {\em Micron},
41(5), 416--422, 2010.
%
\bibitem{dionne1975origin}
GF~Dionne.
\newblock Origin of secondary-electron-emission yield-curve parameters.
\newblock {\em Journal of Applied Physics},46(8), 3347--3351, 1975.
%
\bibitem{henke1979soft}
LB~Henke, J~Liesegang and SD~Smith.
\newblock Soft-x-ray-induced secondary-electron emission from semiconductors and insulators: Models and measurements.
\newblock {\em Physical review B},19(6), 3004, 1979.
%
\bibitem{barut1954mechanism}
AO~Barut.
\newblock The mechanism of secondary electron emission.
\newblock {\em Physical Review},93(5), 981, 1954.
%
\bibitem{kieft2008refinement}
E~Kieft and E~Bosch.
\newblock Refinement of Monte Carlo simulations of electron--specimen interaction in low-voltage SE.
\newblock {\em Journal of Physics D: Applied Physics},
41(21), 215310, 2008.
%
\bibitem{entner2007modeling}
  R~Entner.
  \newblock Modeling and simulation of negative bias temperature instability.
  \newblock {\em 2007.}
 %
 \bibitem{markowichsemiconductor}
PA~Markowich, C~Ringhofer, and C~Schmeiser.
\newblock {\em Semiconductor equations}.
\newblock Springer-Verlag New York, Inc., 1990.
%
\bibitem{raftari2015self}
  Raftari B and Budko NV and Vuik C.
  \newblock Self-consistent drift-diffusion-reaction model for the electron beam interaction with dielectric samples.
  \newblock {\em Journal of Applied Physics}, 118(20):204101, 2015.
%
\bibitem{jerome1985consistency}
Jerome, Joseph W.
\newblock Consistency of semiconductor modeling: an existence/stability analysis for the stationary van Roosbroeck system.
\newblock {\em SIAM journal on applied mathematics}, 45(4): 565--590, 1985.
%
\bibitem{busenberg1993modeling}
Jerome, Joseph W.
\newblock Modeling and analysis of laser-beam-induced current images in semiconductors.
\newblock {\em SIAM journal on applied mathematics}, 53(1): 187--204, 1993.
% 
\bibitem{polak1987semiconductor}
SJ~Polak, C~Den Heijer, WHA~Schilders and P~Markowich.
\newblock Semiconductor device modelling from the numerical point of view.
\newblock {\em International Journal for Numerical Methods in Engineering}, 24(4): 763--838, 1987.
%
\bibitem{ten1993exponential}
JHM~Ten Thije Boonkkamp and WHA~Schilders.
\newblock An exponential fitting scheme for the electrothermal device equations specifically for the simulation of avalanche generation.
\newblock {\em COMPEL-The international journal for computation and mathematics in electrical and electronic engineering},
12(2): 95--111, 1993.
%
\bibitem{dawson1966secondary}
PH~Dawson.
\newblock Secondary electron emission yields of some ceramics.
\newblock {\em Journal of Applied Physics}, 37(9), 3644--3645, 1966.
%
\bibitem{michizono1993dielectric}
S~Michizono, Y~Saito, S~Yamaguchi, S~Anami, N~Matuda and A~Kinbara, A.
\newblock Dielectric materials for use as output window in high-power klystrons.
\newblock {\em IEEE transactions on electrical insulation},
28(4), 692--699, 1993.
%
\bibitem{barnard1997measurements}
J~Barnard, I~Bojko, and N~Hilleret.
\newblock Measurements of the secondary electron emission of some insulators.
\newblock {\em Internal Note (CERN)}, 1997.
%
\bibitem{yong1998determination}
PH~Dawson.
\newblock Determination of secondary electron yield from insulators due to a low-kV electron beam.
\newblock {\em Journal of applied physics}, 84(8), 4543--4548, 1998.
%
\bibitem{kim2010charging}
Kim, Ki Hyun and Akase, Zentaro and Suzuki, Toshiaki and Shindo, Daisuke.
\newblock Charging effects on SEM/SIM contrast of metal/insulator system in various metallic coating conditions.
\newblock {\em Materials transactions},
51(6), 1080--1083, 2010.
%
\bibitem{pickard1970analysis}
Paul S Pickard and Monte V Davis.
\newblock Analysis of electron trapping in alumina using thermally stimulated electrical currents.
\newblock {\em Journal of Applied Physics}, 41(6), 2636--2643, 1970.
%
\bibitem{cazaux1996electron}
J~Cazaux.
\newblock Electron Probe Microanalysis of Insulating Materials: Quantification Problems and Some Possible Solutions.
\newblock {\em X-Ray Spectrometry}, 25(6), 265--280, 1996.
%
\bibitem{dimaria1979radiation}
DJ~DiMaria, LM~Ephrath, and DR~Young.
\newblock Radiation damage in silicon dioxide films exposed to reactive ion etching.
\newblock {\em Journal of Applied Physics}, 50(6), 4015--4021, 1979.
%
\bibitem{dimaria1989trap}
DJ~Dimaria and JW~Stasiak.
\newblock Trap creation in silicon dioxide produced by hot electrons. 
\newblock {\em Journal of Applied Physics}, 65(6), 2342--2356, 1989.
%
\bibitem{buchanan1989coulombic}
DA~Buchanan, MV~Fischetti, and DJ~DiMaria.
Coulombic and neutral trapping centers in silicon dioxide.
\newblock {\em Physical Review B}, 43(2), 1471, 1991.
%
\bibitem{dimaria1991trapping}
DA~Buchanan and and JH~Stathis.
Trapping and trap creation studies on nitrided and reoxidized-nitrided silicon dioxide films on silicon.
\newblock {\em Journal of applied physics}, 70(3), 1500--1509, 1991.
%
  %
\bibitem{fitting1977electron}
H-J Fitting, H~Glaefeke, and W~Wild.
\newblock Electron penetration and energy transfer in solid targets.
\newblock {\em physica status solidi (a)}, 43(1), 185--190, 1977.
 %
 \bibitem{fitting2011secondary}
H-J Fitting and M~Touzin.
\newblock Secondary electron emission and self-consistent charge transport in
  semi-insulating samples.
\newblock {\em Journal of Applied Physics}, 110(4):044111, 2011.
%
 \bibitem{feldman1960range}
  Ch~Feldman.
  \newblock Range of 1-10 keV electrons in solids.
  \newblock {\em Physical Review}, 117(2), 455, 1960.
  %
 \bibitem{kanaya1972penetration}
  Ch~Feldman.
  \newblock Penetration and energy-loss theory of electrons in solid targets. 
  \newblock {\em Journal of Physics D: Applied Physics}, 5(1) 43, 1972.
  %
 \bibitem{lane1954transmission}
  RO~Lane and DJ~Zaffarano.
  \newblock Transmission of 0-40 keV electrons by thin films with application to beta-ray spectroscopy. 
  \newblock {\em Physical Review}, 94(4) 960, 1954.
 % 
 \bibitem{young1956penetration}
  JR~Young.
  \newblock Penetration of electrons in aluminum oxide films. 
  \newblock {\em Physical Review}, 103(2) 292, 1956.
 % 
 \bibitem{yang1987electron}
  KY~Yang and RW~Hoffman.
  \newblock Electron yields and escape depths from Kapton and Teflon. 
  \newblock {\em Surface and Interface Analysis}, 10(2-3) 121--125, 1987.
 % 
 \bibitem{salehi1980experimental}
  M~Salehi and EA~Flinn.
  \newblock An experimental assessment of proposed universal yield curves for secondary electron emission. 
  \newblock {\em Journal of Physics D: Applied Physics}, 13(2) 281, 1980.
 % 
 \bibitem{melchinger1995dynamic}
  Melchinger, A and Hofmann, S.
  \newblock Dynamic double layer model: Description of time dependent charging phenomena in insulators under electron beam irradiation. 
  \newblock {\em Journal of applied physics}, 78(10) 6224--6232, 1995.
 % 
 \bibitem{belhaj2000time}
  M~Belhaj, S~Odof, K~Msellak and O~Jbara.
  \newblock Time-dependent measurement of the trapped charge in electron irradiated insulators: application to Al$_2$O$_3$--sapphire. 
  \newblock {\em Journal of applied physics}, 88(5) 2289--2294, 2000.
 % 
\bibitem{Database}
DC~Joy.
\newblock A database of electron-solid interactions, revision 08-1, 2008.
%
\bibitem{dapor2011secondary}
M~Dapor.
\newblock Secondary electron emission yield calculation performed using two different Monte Carlo strategies.
\newblock {\em Nuclear Instruments and Methods in Physics Research Section B: Beam Interactions with Materials and Atoms},
269(14), 1668--1671, 2011.
%
\bibitem{agarwal1958variation}
BK~Agarwal.
\newblock Variation of secondary emission with primary electron energy.
\newblock {\em Proceedings of the Physical Society}, 71(5), 851, 1958.
%
\bibitem{schreiber2002monte}
E~Schreiber and HJ~Fitting.
\newblock Monte Carlo simulation of secondary electron emission from the insulator SiO$_2$.
\newblock {\em Journal of Electron Spectroscopy and Related Phenomena}, 124(1), 25--37, 2002.
%
\bibitem{cosslett1964multiple}
VE~Cosslett and RN~Thomas.
\newblock Multiple scattering of 5-30 keV electrons in evaporated metal films II: Range-energy relations.
\newblock {\em British Journal of Applied Physics}, 15(11), 1283, 1964.
%
%
\bibitem{fakhfakh2012experimental}
S~Fakhfakh, O~Jbara, S~Rondot, A~Hadjadj and Z~Fakhfakh.
\newblock Experimental characterisation of charge distribution and transport in electron irradiated PMMA.
\newblock {\em Journal of Non-Crystalline Solids},
358(8), 1157--1164, 2012.
%
\bibitem{said2014dependence}
K~Said, G~Damamme, A~Si Ahmed, G~Moya, A~Kallel.
\newblock Dependence of secondary electron emission on surface charging in sapphire and polycrystalline alumina: 
Evaluation of the effective cross sections for recombination and trapping.
\newblock {\em Applied Surface Science},
297, 45--51, 2014.
%
\bibitem{renoud2004secondary}
R~Renoud, F~Mady, C~Attard, J~Bigarre, J-P~Ganachaud.
\newblock Secondary electron emission of an insulating target induced by a well-focused electron beam--Monte Carlo simulation study.
\newblock {\em physica status solidi (a)},
201(9), 2119--2133, 2004.
%
\bibitem{cornet2008electron}
N.~Cornet, D.~G{\oe}uriot, C.~Guerret-Piecourt, D.~Juv{\'e}, D.~Tr{\'e}heux, M.~Touzin, H-J.~Fitting.
\newblock Electron beam charging of insulators with surface layer and leakage currents.
\newblock {\em Journal of Applied Physics},
103(6), 064110, 2008.
%
\bibitem{seiler1983secondary}
H~Seiler.
\newblock Secondary electron emission in the scanning electron microscope.
\newblock {\em Journal of Applied Physics},
54(11), R1--R18, 1983.
%
\bibitem{hughes1979generation}
RC~Hughes.
\newblock Generation, transport, and trapping of excess charge carriers in Czochralski-grown sapphire.
\newblock {\em Physical Review B},
19(10), 5318, 1979.
%
\bibitem{hughes1975hot}
RC~Hughes.
\newblock Hot Electrons in SiO$_2$.
\newblock {\em Physical Review Letters},
35(7), 449, 1975.
%
\bibitem{vaisburd2008poole}
D.I.~Vaisburd,K.E. Evdokimov.
\newblock The Poole--Frenkel effect in a dielectric under nanosecond irradiation by an electron beam with moderate or high current density.
\newblock {\em Russian Physics Journal},
51(12), 1255--1261, 2008.
%
\bibitem{ning1976capture}
TH~Ning.
\newblock Capture cross section and trap concentration of holes in silicon dioxide.
\newblock {\em Journal of Applied Physics},
47(3), 1079--1081, 1976.
%
\bibitem{ning1976high}
TH~Ning.
\newblock High-field capture of electrons by Coulomb-attractive centers in silicon dioxide.
\newblock {\em Journal of Applied Physics},
47(7), 32--3208, 1976.
%
\bibitem{williams1965photoemission}
R~Williams.
\newblock Photoemission of electrons from silicon into silicon dioxide.
\newblock {\em Physical Review},
140(2A), A569, 1965.
%



\end{thebibliography}
\end{document}